\documentclass[12pt]{article}
\pdfoutput=1

\usepackage{jheppub}
\makeatletter
\def\@fpheader{\relax}
\makeatother

\usepackage{amsmath,amssymb}

\title{Residues, modularity, and the Cardy limit of the 4d $\texorpdfstring{\mathcal{N}=4}{N=4}$ superconformal index}

\author[a]{Kevin Goldstein}
\author[a]{\!\!, Vishnu Jejjala}
\author[a,b,c]{\!\!, Yang Lei}
\author[a]{\!\!, Sam van Leuven}
\author[d]{\!\!, Wei Li}

\affiliation[\,a]{Mandelstam Institute for Theoretical Physics, School of Physics, NITheP, and CoE-MaSS, \\ University of the Witwatersrand, 1 Jan Smuts Avenue, Johannesburg, South Africa}
\affiliation[\,b]{The Niels Bohr Institute, Copenhagen University,  Blegdamsvej 17, \\ DK-2100 Copenhagen \O , Denmark}
\affiliation[\,c]{Institute of Theoretical Physics, \\ Chinese Academy of Sciences, 100190 Beijing, P.R.~China}
\affiliation[\,d]{Kavli Institute for Theoretical Sciences (KITS),  \\ University of Chinese Academy of Sciences, 100190 Beijing, P.R.~China}

\emailAdd{kevin.goldstein@wits.ac.za}
\emailAdd{vishnu@neo.phys.wits.ac.za}
\emailAdd{leiyang@ucas.edu.cn}
\emailAdd{svleuven@xs4all.nl}
\emailAdd{weili@mail.itp.ac.cn}

\abstract{
We compute the superconformal index of the $\mathcal{N}=4$ $SU(N)$ Yang-Mills theory through a residue calculation.
The method is similar in spirit to the Bethe Ansatz formalism, except that all poles are explicitly known, and we do not require specialization of any of the chemical potentials.
Our expression for the index allows us to revisit the Cardy limit using modular properties of four-dimensional supersymmetric partition functions.
We find that all residues contribute at leading order in the Cardy limit.
In a specific region of flavour chemical potential space, close to the two unrefined points, in fact all residues contribute universally.
These universal residues precisely agree with the entropy functions of the asymptotically AdS$_5$ black hole and its ``twin saddle'' respectively.
Finally, we discuss how our formula is suited to study the implications of four-dimensional modularity for the index beyond the Cardy limit.
}

\date{}

\begin{document} 

\parskip=10pt

\maketitle 

\section{Introduction}\label{sec:intro}

Recently, there has been renewed interest in the superconformal index of the four-dimensional $\mathcal{N}=4$ $SU(N)$ Yang-Mills theory, which captures the degeneracies (with signs) of the 1/16$^{\mathrm{th}}$ BPS spectrum.
This interest is caused by the fact that, in contrast to the conclusions of an earlier investigation \cite{Kinney:2005ej}, the index is able to reproduce the entropy of supersymmetric black holes with AdS$_5$ asymptotics \cite{Gutowski:2004yv,Chong:2005hr,Chong:2005da,Kunduri:2006ek}.
More precisely, the first series of works \cite{Cabo-Bizet:2018ehj,Choi:2018hmj,Benini:2018ywd} aimed to reproduce a specific ``entropy function'' or free energy from the index, which was previously shown in \cite{Hosseini:2017mds} to lead to the black hole entropy upon a Legendre transformation.
Apart from this success, the index is a rather rich mathematical object and further study has among other things suggested the existence of unknown gravitational saddles \cite{Benini:2018ywd,ArabiArdehali:2019tdm,Cabo-Bizet:2019eaf}.
There are also indications of an interesting structure relating these various saddles, perhaps akin to the $SL(2,\mathbb{Z})$ family of BTZ black holes in three dimensions \cite{Maldacena:1998bw,Dijkgraaf:2000fq,Manschot:2007ha}.

However, the comparison with AdS$_3/$CFT$_2$ is not obvious. 
In particular, this structure arises from modularity of the CFT$_2$ partition function defined on a two-torus.  
The present work was originally geared toward exploring analogous structures in a four-dimensional context.
Specifically, it has been known for a while that certain four-dimensional supersymmetric partition functions, including the superconformal index, have interesting $SL(3,\mathbb{Z})$ properties, as emphasized in \cite{Spiridonov:2012ww,Razamat:2012uv,Brunner:2016nyk}.
These properties can be explained from the fact that the manifolds on which such partition functions are defined can all be viewed as a gluing of two solid three tori along their boundaries, which are identified up to an $SL(3,\mathbb{Z})$ transformation. 
The associated holomorphic block factorization \cite{Peelaers:2014ima,Nieri:2015yia,Longhi:2019hdh} is central to the $SL(3,\mathbb{Z})$ modular property, which was explained in full generality in \cite{Gadde:2020bov}.
However, it turns out that applying the general framework of \cite{Gadde:2020bov} to the problem of interest requires a different approach to the computation of the superconformal index as compared with existing approaches.
The main reason for this is that the previous works, which we briefly summarize below, all lack a certain generality.
This fact prohibits the application of the main modular property of \cite{Gadde:2020bov} to the existing expressions for the index.
In the present work, we will be mainly concerned with the computation of the index through residues and a subsequent Cardy limit via the $SL(3,\mathbb{Z})$ modular property.
We postpone the investigation of the aforementioned structure to a forthcoming publication \cite{Jejjala:2021hlt}.

Let us now turn to a summary of the recent developments. 
In the process, we will also indicate their respective drawbacks for our purposes.
Finally, we will briefly mention how our method is able to avoid these specific drawbacks.
\begin{enumerate}
	\item In \cite{Cabo-Bizet:2018ehj}, the gravitational on-shell action of charged and rotating AdS$_5$ black holes is computed, regularized through a background subtraction scheme.\footnote{See \cite{Cassani:2019mms} for a computation within the framework of holographic renormalization.}
	It was shown that in the supersymmetric limit, this gravitational on-shell action precisely reproduces the entropy function of \cite{Hosseini:2017mds}, which upon Legendre transformation yields the black hole entropy.
	This gravity computation clarifies the role of the complex chemical potentials and the (complex) constraint they are subjected to, first noticed in \cite{Hosseini:2017mds}, as arising from regularity of a Killing spinor in the black hole background; it also dictated the boundary background on which the authors of \cite{Cabo-Bizet:2018ehj} then computed the field theory partition function via a supersymmetric localization computation. 	
	The resulting partition function has the form $Z=e^{-\mathcal{F}}\mathcal{I}$, with $\mathcal{I}$ the superconformal index and $\mathcal{F}$ a generalized supersymmetric Casimir energy,\footnote{See \cite{Assel:2014paa,Assel:2015nca} for the definition of the ordinary supersymmetric Casimir energy.} which turns out to equal \emph{minus} the on-shell action.\footnote{In the large $N$ or Cardy limit (and at the relevant \emph{complex} values of the chemical potentials), $\log \mathcal{I}$ reproduces the entropy function  \cite{Choi:2018hmj,Benini:2018ywd}. 
	This indicates that the bulk computation corresponds to a scheme in the field theory where the path integral gives rise to a partition function of the form $Z=\mathcal{I}$, i.e. without the Casimir energy prefactor \cite{Cabo-Bizet:2018ehj}.} 
	The appearance of a Casimir like energy in the computation of black hole entropy is reminiscent of AdS$_3$/CFT$_2$, where modularity in the field theory connects the Casimir energy with the entropy of black hole states. 
	We will see in Section \ref{sec:cardy-limit-index} that indeed a modular property of the superconformal index precisely yields this generalized Casimir energy, thereby confirming this analogy in a more precise manner (see also \cite{Gadde:2020bov}).

	From our perspective, the main disadvantage to this approach is the fact that the gravitational solution plays a crucial role in the analysis.
	This makes it difficult to see how this method will lead to insight into unknown gravitational solutions.
		
	\item Various authors have considered a Cardy-like limit of the partition function \cite{Choi:2018hmj,Honda:2019cio,ArabiArdehali:2019tdm,Kim:2019yrz,Cabo-Bizet:2019eaf,ArabiArdehali:2019orz,GonzalezLezcano:2020yeb,David:2020ems}.
	This limit sends the chemical potentials $\tau$ and $\sigma$, associated with the angular momenta, to zero.
	It is analogous to the high temperature limit of the torus partition function in two-dimensional CFT, hence the name.
	In this limit, the computation of the index simplifies significantly, although one still has to rely on a saddle point approximation of the gauge integral to obtain the final result.
	It was pointed out in \cite{Gadde:2020bov} that there is an interesting connection between this Cardy limit and a certain modular property of the index.
	However, this modular property cannot be used for all values of the gauge parameters, and therefore its use has to be justified a posteriori by the precise value of the saddle.	
	We believe that the implications of modularity will be more transparent by first computing the gauge integral (exactly) and only then applying the modular property.
	In this way, one can a priori justify its use and the effect of the Cardy limit is not obscured by a subsequent saddle point approximation.
	\item The Bethe Ansatz method \cite{Closset:2017bse,Closset:2018ghr, Benini:2018mlo} allows one to compute the gauge integral exactly and can be used to evaluate the large-$N$ limit of the index \cite{Benini:2018ywd,Benini:2020gjh}.
	Using this method, the index can be expressed as a sum over ``Bethe vacua'' and is reminiscent of a sum over saddles.
	The Bethe Ansatz method was the first to point out a rich phase structure exhibited by the large-$N$ limit of the index, by showing that in different regimes of parameter space different Bethe vacua provide the dominant contribution to the index.
	However, there are some unsatisfactory aspects such as the fact that not all Bethe vacua are known (see however \cite{ArabiArdehali:2019orz,GonzalezLezcano:2020yeb}).
	Also, one requires a certain specialization of the $\tau$ and $\sigma$ chemical potentials.
	More precisely, the initial work \cite{Benini:2018ywd} required $\tau=\sigma$, which was later generalized to $\tau/\sigma\in \mathbb{Q}$ \cite{Benini:2020gjh}.
	But even for the more general setting, the modular property of \cite{Gadde:2020bov} does not hold.\footnote{However, a version of the modular property exists for this specialization, see Theorem 5.2 in \cite{Felder_2000}.
	This property was in fact employed in \cite{Benini:2018ywd,Benini:2020gjh} to compute the large-$N$ limit of the index.}
	\item Another line of work has attempted a saddle point approximation of the gauge integral in the large-$N$ limit, as opposed to the Cardy limit \cite{Cabo-Bizet:2019eaf,Cabo-Bizet:2020nkr}.\footnote{See also \cite{Copetti:2020dil} for a different approach, based on a truncated version of the $\mathcal{N}=4$ matrix model.}
	Due to an elliptic extension of the gauge parameters, it is possible to find a large class of saddles with interesting interrelations.
	In addition, this analysis makes it possible to study the phase structure of the 1/16$\mathrm{th}$ BPS sector and predicts the existence of new gravitational saddles.
	The main drawback of this approach is the apparent lack of an a priori justification of the elliptic extension.
	Indeed, such ellipticity is naturally associated with the presence of large gauge transformations along two non-trivial one-cycles in the geometry, whereas $S^3\times S^1$ only has a single non-trivial one-cycle.
	In addition, this approach so far has required the specialization $\tau=\sigma$, which as we already mentioned is problematic for the use of the full modular property.
	\item Most recently, there have also been numerical efforts to study the superconformal index \cite{Murthy:2020rbd,Agarwal:2020zwm}. These methods both require specializations of the parameters and truncations, which obscure modular properties.
\end{enumerate}
The main issues we have pointed out with the existing methods can be summarized as: specialization of parameters and/or working at the level of the gauge integrand.
We obtain another, exact representation of the index similar in structure to the Bethe Ansatz analysis.
In particular, we first evaluate the gauge integral through residues and therefore do not rely on saddle point approximations.
In addition, we do not require any specialization of the parameters to perform the computation.
However, as one varies the chemical potentials some poles may enter or exit the contour of integration.
Although this prohibits us from finding a fully explicit expression for the index for general parameters, this is inconsequential for the study of the Cardy limit.
The residue analysis is the same as the one that relates the Higgs branch localization formulas of $\mathcal{N}=1$ gauge theories with fundamental matter to the gauge integral expression of the index \cite{Yoshida:2014qwa,Peelaers:2014ima,Nieri:2015yia}.
The term ``Higgs branch localization'' in the case of the $\mathcal{N}=4$ is inappropriate, since the $\mathcal{N}=4$ theory does not have a Higgs branch.
We will therefore refrain from using this terminology, even though the analysis and expressions will be completely analogous to those obtained in gauge theories using Higgs branch localization.

The computation of the index is performed in Section \ref{sec:comp-n=4-index}, first for $SU(2)$ gauge group in Section \ref{ssec:su2-index} and then for $SU(N)$ in Section \ref{ssec:suN-index}.
Only after this step will we take the Cardy limit of the resulting expression in Section \ref{sec:cardy-limit-index}, using a modular property of the index.
We conclude and discuss our results in Section \ref{sec:conc}.
In Appendix \ref{app:props-theta-gamma} we collect definitions and some useful properties of special functions that appear in the index. 
In Appendix \ref{app:unrefined-limit-index}, we compute the unrefined limit of the index, which requires special care, and its Cardy limit.
Finally, we collect the expression for the anomaly polynomial in Appendix \ref{app:anomaly-pol} and briefly discuss how it is related to the standard expression of the entropy function used in the main text.

\section{Computation of the \texorpdfstring{$\mathcal{N}=4$}{N=4} index}\label{sec:comp-n=4-index}

In this section, we will perform the computation of the superconformal index for the $\mathcal{N}=4$ $SU(N)$ Yang-Mills theory.
We will warm-up with the $SU(2)$ case and then proceed to general $N$.
Before turning to the computation, we will describe the index in some detail and in the process set up notation.
See \cite{Gadde:2020yah} for a recent review.

The superconformal index can be expressed as follows:
\begin{equation}
I(\phi_{1,2};\tau,\sigma)=\mathrm{tr}_{\mathcal{H}_{Q}}(-1)^F p^{J_1}q^{J_2}(pq)^{\frac{r_1-r_3}{2}}f_1^{r_2+r_3}f_2^{r_3}.
\end{equation}
Here, 
\begin{equation}
p=e^{2\pi i\sigma},\quad q=e^{2\pi i \tau}, \quad f_1=e^{2\pi i\phi_1},\quad f_2=e^{2\pi i \phi_2}.
\end{equation}
Our parametrization is equivalent to the one used in \cite{Benini:2018ywd} upon identifying $f_i=y_i$.
Furthermore, $\mathcal{H}_Q$ is the 1/16$^{\mathrm{th}}$ BPS Hilbert space corresponding to those states on which:
\begin{equation}
\lbrace Q,Q^\dagger\rbrace=E-J_1-J_2-\tfrac{3}{2}(r_1+\tfrac{2}{3}r_2+\tfrac{1}{3}r_3)
\end{equation}
vanishes.
The charges $J_{1,2}$ are the rotation generators along the Euler angles of $S^3$ and the $r_i$ correspond to the Cartan generators of $SU(4)$.
Thinking of the $\mathcal{N}=4$ theory in an $\mathcal{N}=1$ language, the R-symmetry charges of the three chiral multiplets are given by:
\begin{equation}
(r_1,r_2,r_3)=(0,1,0),(1,-1,1),(1,0,-1).
\end{equation}
The $SO(6)$ Cartan generators $R_i$ used in \cite{Benini:2018ywd} are related to the $r_i$ via:
\begin{equation}
R_1=r_1+2r_2+r_3, \quad R_2=r_1+r_3,\quad R_3=r_1-r_3.
\end{equation}
Each $R_i$ equals 2 for a single chiral multiplet and vanishes on the other two.
Finally, the $\mathcal{N}=1$ superconformal $U(1)$ R-charge $r$ corresponds to:
\begin{equation}
r=r_1+\tfrac{2}{3}r_2+\tfrac{1}{3}r_3
\end{equation}

Since the index is independent of continuous couplings, one can compute it at weak coupling.
In this case, the trace can  be explicitly performed and the resulting expression is given by:
\begin{equation}\label{eq:defn-suN-index}
I_N=\frac{\kappa_N}{N!} \prod^{N-1}_{k=1}\oint_{\left|x_k\right|=1}\frac{dx_k}{2\pi i x_k}\prod_{1\leq i\neq j\leq N}\frac{\prod^3_{a=1}\Gamma(x_{ij}f_a)}{\Gamma(x_{ij})}.
\end{equation}
The integral over the gauge fugacities $x_i=e^{2\pi i u_i}$ ($x_{ij}=x_ix_j^{-1}$) ensures the projection onto gauge invariant states.
Moreover, for notational convenience we have written the elliptic $\Gamma$ function $\Gamma(u;\tau,\sigma)$ as a function of fugacities:
\begin{align}\label{eq:defn-ell-gamma}
\begin{split}
\Gamma(x)\equiv\Gamma(u;\tau,\sigma)=\prod^{\infty}_{m,n=0}\frac{1-x^{-1}p^{m+1}q^{n+1}}{1-x p^{m}q^{n}}&=\mathrm{P.E.}\left(\frac{x-x^{-1}pq}{(1-p)(1-q)}\right)\\
&\equiv\exp\left(\sum^{\infty}_{l=1}\frac{1}{l}\frac{x^l-(x^{-1}pq)^l}{(1-p^l)(1-q^l)}\right),
\end{split}
\end{align}
where the second equation indicates how the elliptic $\Gamma$ function can be thought of as a generating function for multiletter indices from a single letter index.
More precisely, writing the integrand of \eqref{eq:defn-suN-index} in terms of plethystic exponentials, the argument of the combined exponentials is $1-f$ with $f$ the 1/16$^{\mathrm{th}}$ single letter index of the $\mathcal{N}=4$ theory. 
The additional 1 originates from the Vandermonde determinant obtained by replacing the matrix integral with an integral over eigenvalues \cite{Aharony:2003sx,Kinney:2005ej}.

In addition, we defined $f_3=pqf_1^{-1}f_{2}^{-1}$ and $x_N=(x_1\cdots x_{N-1})^{-1}$, the latter corresponding to the $SU(N)$ constraint.
Furthermore, $\kappa_N$ consists of the Cartan factors of both the chiral multiplets and the vector multiplet, and is given by:
\begin{equation}
\kappa_N=(p;p)^{N-1}_{\infty}(q;q)^{N-1}_{\infty}\left(\Gamma(f_1)\Gamma(f_2)\Gamma(f_3)\right)^{N-1},
\end{equation}
where the $q$-Pochhammer symbol is defined in \eqref{eq:defn-qpoch}.

Finally, for convergence of the product formula for the elliptic $\Gamma$ function, one should require $|p|,|q|<1$.
For the summation formula, one needs in addition:
\begin{equation}\label{eq:domain-of-conv}
|pq|<f_a<1.
\end{equation}
Notice that if the second requirement holds for $f_{1,2}$, it automatically holds for:
\begin{equation}
f_3=pq(f_1f_2)^{-1}.
\end{equation}
The domain of convergence of the $\Gamma$ functions can be extended outside the unit disk.
This is discussed in more detail in Appendix \ref{app:props-theta-gamma}.

\subsection{\texorpdfstring{$SU(2)$}{SU(2)} index}\label{ssec:su2-index}

For $SU(2)$ gauge group, the gauge integral consists of a single contour integral:
\begin{equation}
I_2=\frac{\kappa_2}{2}\oint_{\left|x\right|=1} \frac{dx}{2\pi i x}\frac{\prod_{b=1}^3\Gamma(x^{\pm 2}f_b)}{\Gamma(x^{\pm 2})},
\end{equation}
where we defined $\Gamma(x^{\pm})\equiv \Gamma(x)\Gamma(x^{-1})$.
The computation of this integral by residues has already been done in \cite{Peelaers:2014ima}.
More precisely, in that paper the index is computed for an $\mathcal{N}=1$ $SU(2)$ vector multiplet coupled to $N_f$ fundamental chiral and $N_{\bar{f}}$ anti-chiral multiplets.  
Even though we consider adjoint chiral multiplets, for $SU(2)$ gauge group the only distinction at the level of the index is the power $x^2$ instead of $x$ appearing in the argument of the $\Gamma$ functions.
Keeping this minor distinction in mind, the following computation may be viewed as a special case of the computation in \cite{Peelaers:2014ima} for $N_f=3$ chiral multiplets and $N_{\bar{f}}=0$ anti-chiral multiplets.

We will compute the integral by picking up residues from the poles of the integrand inside the unit circle.
From its definition \eqref{eq:defn-ell-gamma}, it is not difficult to see that the $\Gamma(x)$ function has simple poles at $x=p^{-k}q^{-l}$ for $k,l\in\mathbb{Z}_{\geq 0}$.
Therefore, each $\Gamma(x^{2}f_a)$ in the numerator has poles at:
\begin{equation}
x^{ 2}=f_a^{-1}p^{-k}q^{-l},
\end{equation}
whereas each $\Gamma(x^{-2}f_a)$ factor has poles at:
\begin{equation}\label{eq:su2-pole-pq}
x^{2}=f_ap^{k}q^{l}.
\end{equation}
Given our restriction \eqref{eq:domain-of-conv}, only the latter poles lie inside the unit circle and therefore only these will contribute to the residue sum.

The $\Gamma$ functions in the denominator do not contribute poles.
One way to see this is to note that:
\begin{equation}\label{eq:gamma-theta-id}
\frac{1}{\Gamma(x)\Gamma(x^{-1})}=\theta_p(x)\theta_q(x^{-1}),
\end{equation}
where the $\theta_q$ function is defined in \eqref{eq:defn-theta}.
Now, the statement that the $\Gamma$ functions in the denominator do not contribute poles follows directly from the fact that the right hand side of \eqref{eq:gamma-theta-id} only has zeros.

For future use, we will now derive the following residue corresponding to a basic pole ($k=l=0$) of the elliptic $\Gamma$ function: 
\begin{align}\label{eq:basic-su2-res}
\begin{split}
\mathrm{Res}_{x=\pm f^{\frac{1}{2}}}\left(\frac{\Gamma(x^{-2}f)}{x} \right)&=\oint_{\gamma(\pm f^{\frac{1}{2}})} \frac{dx}{2\pi ix} \frac{1}{(1-x^{-2}f)}\prod_{m,n\geq 0}\frac{1-x^{2}f^{-1}p^{m+1}q^{n+1}}{1-x^{-2}fp^{m+1}q^{n+1}}\\
&\times \prod_{m\geq 0}\frac{1}{1-x^{-2}fp^{m+1}}\frac{1}{1-x^{-2}fq^{m+1}}=\frac{1}{2(p;p)_{\infty}(q;q)_{\infty}},
\end{split}
\end{align}
where the contour $\gamma(\pm f^{\frac{1}{2}})$ is an infinitesimal circle around $x=\pm f^{\frac{1}{2}}$.

To compute the residues at the more general poles \eqref{eq:su2-pole-pq}, we need the following two properties of the $\Gamma$ function (see Appendix \ref{app:props-theta-gamma} for more details):
\begin{align}\label{eq:basic-shift-props-Gamma}
\begin{split}
\Gamma(p^{k}q^{l}x)&=\Gamma(x)\left(-xp^{\frac{k-1}{2}}q^{\frac{l-1}{2}}\right)^{-kl}\prod^{k-1}_{m=0}\theta_q(xp^{m})\prod^{l-1}_{n=0}\theta_p(xq^{n}),\\
\Gamma(p^{-k}q^{-l}x)&=\frac{\Gamma(x)}{\left(-x^{-1}p^{\frac{k+1}{2}}q^{\frac{l+1}{2}}\right)^{-kl}\prod^{k}_{m=1}\theta_q(xp^{-m})\prod^{l}_{n=1}\theta_p(xq^{-n})}.
\end{split}
\end{align}

Now we are ready to evaluate the contour integral. 
We deform the contour such that it splits into a sum of three ``towers'' of contours, encircling the poles at $x^2=f_ap^kq^l$ for $a=1,2,3$ and $k,l\geq 0$.
The sum of the residues of the integrand at the two basic poles $x^2=f_a$ reads:
\begin{equation}
\frac{\prod^3_{b=1}\Gamma(f_bf_a)\prod_{b\neq a}\Gamma(f_bf_a^{-1})}{\Gamma(f_a)\Gamma(f_a^{-1})}\frac{1}{(p;p)_{\infty}(q;q)_{\infty}}.
\end{equation}
On a general pole $x^2=f_ap^kq^l$, instead we have:\footnote{We use the notation $x^2=f_ap^kq^l$ to denote the collection of the two poles $x=\pm(f_ap^kq^l)^{\frac{1}{2}}$. This is also the reason why in the expression for the residue, the factor of $\tfrac{1}{2}$ has disappeared with respect to \eqref{eq:basic-su2-res}.}
\begin{align}
\begin{split}
&\frac{\prod^3_{b=1}\Gamma(f_bf_ap^kq^l)\prod_{b\neq a}\Gamma(f_bf_a^{-1}p^{-k}q^{-l})}{\Gamma(f_ap^kq^l)\Gamma(f_a^{-1}p^{-k}q^{-l})}\times \mathrm{Res}_{x^{2}=f_ap^kq^l}\left(\frac{\Gamma(x^{-2}f_a)}{x}\right)\\
&=\frac{\prod^3_{b=1}\Gamma(f_bf_a)\prod_{b\neq a}\Gamma(f_bf_a^{-1})}{\Gamma(f_a)\Gamma(f_a^{-1})}(pq)^{-kl}\frac{\prod^k_{m=1}\theta_q(f_a^{-1}p^{-m})\prod^l_{n=1}\theta_p(f_a^{-1}q^{-n})}{\prod^{k-1}_{m=0}\theta_q(f_ap^{m})\prod^{l-1}_{n=0}\theta_p(f_aq^{n})}\\
&\times \prod^3_{b=1}(f_b^2q^{-1}p^{-1})^{-kl}\frac{\prod^{k-1}_{m=0}\theta_q(f_bf_ap^{m})\prod^{l-1}_{n=0}\theta_p(f_bf_aq^{n})}{\prod^{k}_{m=1}\theta_q(f_bf_a^{-1}p^{-m})\prod^{l}_{n=1}\theta_p(f_bf_a^{-1}q^{-n})}\frac{1}{(p;p)_{\infty}(q;q)_{\infty}}.
\end{split}
\end{align}
Here, we used the shift properties of the $\Gamma$ function listed above to extract the same prefactor as on the basic pole.
Noting that $f_1f_2f_3p^{-1}q^{-1}=1$ and that the $\frac{1}{(p;p)_{\infty}(q;q)_{\infty}}$ cancels that same part in $\kappa_2$, upon summing all residues one arrives at the final result:
\begin{align}
\begin{split}
&I_2=\frac{\Gamma(f_1)\Gamma(f_2)\Gamma(f_3)}{2}\sum^3_{a=1}\frac{\prod^3_{b=1}\Gamma(f_bf_a)\prod_{b\neq a}\Gamma(f_bf_a^{-1})}{\Gamma(f_a)\Gamma(f_a^{-1})}\times\\
& \sum_{k,l\geq 0}\Bigg[\frac{\prod^k_{m=1}\theta_q(f_a^{-1}p^{-m})\prod^l_{n=1}\theta_p(f_a^{-1}q^{-n})}{\prod^{k-1}_{m=0}\theta_q(f_ap^{m})\prod^{l-1}_{n=0}\theta_p(f_aq^{n})}\prod^3_{b=1}\frac{\prod^{k-1}_{m=0}\theta_q(f_bf_ap^{m})\prod^{l-1}_{n=0}\theta_p(f_bf_aq^{n})}{\prod^{k}_{m=1}\theta_q(f_bf_a^{-1}p^{-m})\prod^{l}_{n=1}\theta_p(f_bf_a^{-1}q^{-n})}\Bigg].
\end{split}
\end{align} 
Let us note here that the sum over $k,l$ factorizes into a part that only contains $\theta_{q}$ functions and a part that only contains $\theta_{p}$ functions.
In the Higgs branch localization literature, these functions are called vortex partition functions.
As already mentioned in the introduction, the $\mathcal{N}=4$ theory strictly speaking does not have a Higgs branch.
Despite this, we will still use the terminology of vortex partition functions to indicate these products of $\theta$ functions because of their similarity to the vortex partition functions of gauge theories with fundamental matter.
The final result can then be expressed as:
\begin{align}\label{eq:su2-index-final}
\begin{split}
&I_2=\frac{\Gamma(f_1)\Gamma(f_2)\Gamma(f_3)}{2}\sum^3_{a=1}\frac{\prod^3_{b=1}\Gamma(f_bf_a)\prod_{b\neq a}\Gamma(f_bf_a^{-1})}{\Gamma(f_a)\Gamma(f_a^{-1})}Z_V(\phi_a,\sigma;\tau)Z_V(\phi_a,\tau;\sigma),
\end{split}
\end{align} 
where:
\begin{align}\label{eq:vortex-factors}
\begin{split}
Z_{V}(\phi,\sigma;\tau)= \sum_{k\geq 0}\frac{\prod^k_{m=1}\theta_q(f_a^{-1}p^{-m})}{\prod^{k-1}_{m=0}\theta_q(f_ap^{m})}\prod^3_{b=1}\frac{\prod^{k-1}_{m=0}\theta_q(f_bf_ap^{m})}{\prod^k_{m=1}\theta_q(f_bf_a^{-1}p^{-m})} .
\end{split}
\end{align} 

Let us make some final comments.
First of all, notice that this computation is valid for values of the chemical potentials obeying \eqref{eq:domain-of-conv}.
In particular, we do not need to constrain the values of $\tau$ and $\sigma$, which is required in the Bethe Ansatz formalism \cite{Benini:2018mlo}.
Secondly, the unrefined limit $f_1=f_2=f_3=(pq)^{\frac{1}{3}}$ of our expression \eqref{eq:su2-index-final} is singular.
This can be traced to the fact that in this limit, the integrand develops cubic instead of simple poles.
Therefore, to access the unrefined limit in our formalism one has to redo the residue computation, now taking into account the higher order poles. 
We defer this analysis to Appendix \ref{app:unrefined-limit-index}.

\subsection{\texorpdfstring{$SU(N)$}{SU(N)} index}\label{ssec:suN-index}

We would now like to do a similar computation for $SU(N)$ gauge group.
The expression for the index was given in \eqref{eq:defn-suN-index}.
For convenience, we implement the $SU(N)$ constraint such that the index can be written as:
\begin{align}\label{eq:sci-suN}
\begin{split}
I_N=\frac{\kappa_N}{N!}\prod^{N-1}_{k=1}\oint_{\left|x_k\right|=1}\frac{dx_k}{2\pi i x_k}\prod^{N-1}_{i< j}\frac{\prod^3_{b=1}\Gamma(x_{ij}^{\pm}f_b)}{\Gamma(x_{ij}^{\pm})}\prod^{N-1}_{i=1}
\frac{\prod^3_{b=1}\Gamma((x_1 \cdots x_i^2\cdots x_{N-1})^{\pm }f_b)}{\Gamma((x_1 \cdots x_i^2\cdots x_{N-1})^{\pm })}.
\end{split}
\end{align}
To compute such multidimensional contour integrals, one cannot in general  resort to Cauchy's theorem directly.
The reason for this is that poles may not factorize in their dependence on $x_i$, as is indeed the case for the integrand at hand.
Let us therefore briefly review how to deal with such multivariate residue integrals.

\paragraph{Interlude on multivariate residues:}

Let $g(x) = \big(g_1(x), \ldots, g_n(x) \big) : \mathbb{C}^n \to \mathbb{C}^n$ and $h : \mathbb{C}^n \to \mathbb{C}$ be holomorphic functions.
We are interested in computing the residue of the meromorphic $n$-form $\omega$:
\begin{align}
\omega = \frac{h(x) d x_1\cdots d x_n}{g_1(x)\cdots g_n(x)}.
\end{align}
A pole of $\omega$ is defined as an isolated point $p \in \mathbb{C}^n$
such that $g(p) = 0$.
The residue of $\omega$ is now computed by the integral:
\begin{align}
\mathrm{Res}_{x=p}(\omega) =
\frac{1}{(2\pi i)^n}\oint_{\gamma_p}
\frac{h(x) d x_1\cdots d x_n}{g_1(x)\cdots g_n(x)},
\end{align}
where $\gamma_p$ is an $n$-torus centered around $p$.

One can evaluate the following Jacobian determinant at a pole $x=p$:
\begin{align}
J_p\equiv \det\left(  \frac{\partial g_i}{\partial x_j}  \right)\bigg|_{x = p}.
\end{align}
If $J_p \neq 0$, which will turn out to be true away from the unrefined limit,\footnote{A more general formula for the residue in the case of $J_p=0$ is given in Appendix \ref{app:suN-index}.} one can perform the coordinate transformation $y_i = g_i(x)$ such that the poles factorize in $y$ coordinates.
Then, the residue can be evaluated as a product of one-dimensional residue integrals:
\begin{align}\label{eq:non-deg-res}
\mathrm{Res}_{x=p}(\omega)
= \frac{1}{(2\pi i)^n}
\oint_{\gamma_p}
\frac{h\big(g^{-1}(y)\big) d y_1\cdots d y_n}{J_p y_1 \cdots y_n}= \frac{h(p)}{J_p}.
\end{align}
We will apply this general formula to the computation of \eqref{eq:sci-suN} by first classifying all the poles of the integrand.
Subsequently, we deform the contour such that it splits into a sum of $(N-1)$-tori, each of which encircles a pole of the integrand $p=(x_1,\ldots,x_{N-1})$ for which all $x_i$ lie inside $|x_i|=1$. 
Other poles will not contribute to the resulting residue sum.

\paragraph{Back to the index:}

For the same reason as in the $SU(2)$ case, all poles of the integrand originate from the $\Gamma$ functions in the numerator.
The total number of $\Gamma$ functions in the numerator is equal to $3(N^2-N)$.
A (simple) pole of the integrand is realized at those points where $N-1$ of these $\Gamma$ functions have a pole.
Therefore, poles of the integrand can be found by selecting $N-1$ $\Gamma$ functions in the numerator $\Gamma(y_i)$ and subsequently solving the system of equations $y_i=p^{-k_i}q^{-l_i}$ for some $k_i,l_i\geq 0$.

The $3(N^2-N)$ pole equations are linear equations when written in terms of the chemical potentials:
\begin{align}\label{eq:lin-eqns-set}
\begin{split}
1&: u_i-u_j=\phi_{a_{ij}}+k_{ij}\sigma+l_{ij}\tau, \quad 1\leq i\neq j\leq N-1,\; a_{ij}=1,2,3,\\
2&:u_1+\ldots  +2u_i+\ldots+u_{N-1}=\mp (\phi_{a_i}+k_{i}\sigma+l_{i}\tau), \quad i=1,\ldots,N-1,\; a_i=1,2,3,
\end{split}
\end{align}
for some $k_{ij},l_{ij},k_i,l_i\geq 0$.
Selecting $N-1$ of these equations and solving them for $u_i$ leads to a pole of the integrand.
At first sight, it may seem that for large-$N$ this leads to a huge number of poles to analyze.
However, there turns out to be a significant reduction in the number of poles that contribute non-trivially to the full residue sum as we will now argue.

First of all, for the system of $N-1$ equations to be solvable, at least one of the equations has to be of the type in the second line of \eqref{eq:lin-eqns-set}.
In particular, the system is solvable when choosing the $N-1$ equations to be all of the second type.
More generally, given some number of equations of the second type, the system remains solvable as long as one adds equations of the first line such that the system can be rewritten in the form of $N-1$ equations of the second type.
For example, say we choose the $i^{\mathrm{th}}$ equation from the second line, while the rest of the equations come from the first line.
Then, the system is solvable when we choose the $N-2$ $u_i-u_j$ equations for $1\leq j\leq N-1$ and $j\neq i$, since subtracting these equations from the  $i^{\mathrm{th}}$ equation of the second line brings us back to the system with $N-1$ equations of the second type.

Now comes the first main simplification of the analysis.
Suppose that at least one of the equations in the (solvable) system is of the first type.
For definiteness, let this equation be:
\begin{equation}
u_i-u_j=\phi_{a_{ij}}+k_{ij}\sigma+l_{ij}\tau.
\end{equation}
For each such equation of the first type in the system, there exists another system of $N-1$ equations where in all equations the labels $i$ and $j$ are exchanged and such that $\phi_{a_{ij}}=\phi_{a_{ji}}$, $k_{ij}=k_{ji}$ and $l_{ij}=l_{ji}$.
The solutions to these two systems are identical, up to an exchange of $u_i$ and $u_j$.
Together with the fact that the residue formula \eqref{eq:non-deg-res} is odd under the exchange of two integration variables $x_i$ and $x_j$, this implies that the residues corresponding to these two systems are equal but of opposite sign.
Therefore, they will cancel in the sum over residues.

This leaves us with the analysis of the class of unpaired poles: the solution to the system of $N-1$ equations of the second type.
Let us consider first the $N-1$ equations with the $+$ sign on the right hand side, since these poles have the best chance of lying inside the unit circles.
The solution of this system is as follows:
\begin{equation}\label{eq:pole-suN}
p^{(n)}: \qquad x_i^N=\frac{f_{a_i}^{N}p^{Nk_i-\sum_{j}k_j}q^{Nl_i-\sum_{j}l_j}}{\prod_{j}f_{a_j}}.
\end{equation}
Here, we indicate by $p^{(n)}$ with $n=1,\ldots,N$, the $N$ distinct solutions for the $x_i$.

If one wishes to replace the $+$ sign in the $i^{\mathrm{th}}$ equation with a $-$ sign, the solution can be obtained from \eqref{eq:pole-suN} by taking:
\begin{equation}\label{eq:plus-to-min}
f_{a_i}\to f_{a_i}^{-1},\quad (k_i,l_i)\to -(k_i,l_i). 
\end{equation}
Let us now check whether the pole \eqref{eq:pole-suN} indeed lies within all unit circles.
First of all, it is easy to see that a pole with all $k_i$ equal and all $l_i$ equal lies inside all unit circles for comparable values of the $f_{a}$.
This is because in this case the net power of fugacities is positive for every $i$.
However, we will quickly see that poles with all $k_i$ equal or all $l_i$ equal have a vanishing residue.
In fact, for the residue to be non-vanishing all but a few $k_i$ have to be distinct and similarly for the $l_i$.
A minimal choice of a pole with non-vanishing residue would for example be:
\begin{equation}
k_i=i-1, \quad l_i=N-1-i.
\end{equation}
This specific choice leads to the following absolute value:
\begin{equation}\label{eq:pole-suN-spec}
|x_i^N|= \left|\frac{f_{a_i}^{N}p^{N(i+\frac{1}{2})-\frac{1}{2}N^2-1}q^{N(i+\frac{1}{2})+\frac{1}{2}N^2-1}}{\prod_{j}f_{a_j}}\right|.
\end{equation}
One may convince oneself that there is a finite domain in parameter space where all $x_i$ are inside their unit circles.
For example, when $|p|\approx |q|$, one finds that:
\begin{equation}
|x_i^N|= \left|\frac{f_{a_i}^{N}p^{N-2}}{\prod_{j}f_{a_j}}\right|.
\end{equation}
For comparable values of the $|f_a|$, this point will lie inside all unit circles, while at large-$N$ the point lies inside all unit circles for all values of the $|f_a|$.

Furthermore, the domain for which \eqref{eq:pole-suN-spec} lies inside all unit circles can also be made parametrically large by considering the following shift:
\begin{equation}
(k_i,l_i)\to (k_i+n_1,l_i+n_2), \quad n_1,n_2>0.
\end{equation}
This new pole will have an additional factor of $p^{n_1}q^{n_2}$ in the numerator, thus enlarging the domain in parameter space for which the pole lies inside all unit circles. 
For large enough $n_{1,2}$, one does not need to take $|p|\approx |q|$ for this to be true.

A complete analysis of all poles lying inside the unit circles, including systems of equations with some $+$ signs are replaced with $-$ signs, is beyond the scope of this paper. 
In fact, for our purposes, i.e.\ computing the Cardy limit of the index using a modular property, it will be sufficient to know that there is at least one pole with non-vanishing residue.
As argued above this is always the case, irrespective of where we are in parameter space. 

We will now continue to compute the residues for the poles where all $N-1$ equations are taken with a $+$ sign.
Other poles originating from a set of equations including $-$ signs can be obtained from these residues through the transformation \eqref{eq:plus-to-min}.

The computation of the residue makes use of the formula \eqref{eq:non-deg-res}.
So, we first have to evaluate the Jacobian of the pole \eqref{eq:pole-suN}.
For the functions $g_i(x)$, we take:
\begin{equation}
g_i(x)=x_i\left(1-\frac{f_{a_i}p^{k_i}q^{l_i}}{x_1\cdots x_i^2\cdots x_{N-1}}\right).
\end{equation}
Then, for each of the $N$ poles \eqref{eq:pole-suN} the Jacobian consists of:
\begin{equation}
\left(  \frac{\partial g_i}{\partial x_j}  \right)\bigg|_{x = p^{(n)}}=\begin{cases}
2,&  j=i,\\
\frac{f_{a_i}p^{k_i}q^{l_i}}{f_{a_j}p^{k_j}q^{l_j}},& j\neq i.
\end{cases}
\end{equation}
It is not difficult to check that this implies:
\begin{equation}
J_{p^{(n)}}=\det\left(  \frac{\partial g_i}{\partial x_j}  \right)\bigg|_{x = p^{(n)}}=N.
\end{equation}
Notice that the Jacobian is independent of $n$.
In addition, the function $h(x)$ in our case can also be seen to be equal on each of the $N$ poles in \eqref{eq:pole-suN}.
Therefore, upon summing the contributions of the $N$ poles, the residue formula:
\begin{align}
\mathrm{Res}_{\sum p^{(n)}}(\omega)\equiv \sum^N_{n=1}\frac{h(p^{(n)})}{N}
= h(p^{(m)}),
\end{align}
has a trivial contribution from the Jacobian and one just has to evaluate $h(x)$ on any of the $N$ poles $p^{(m)}$.

To finish the computation of the residue, we first note that on any of the $N$ poles \eqref{eq:pole-suN}:
\begin{equation}
x_{ij}=f_{a_i}f_{a_j}^{-1}p^{k_i-k_j}q^{l_i-l_j}.
\end{equation}
Now we can straightforwardly evaluate the residue:
\begin{align}\label{eq:residue-basic-pole-suN}
\begin{split}
&\mathrm{Res}_{\sum p^{(n)}}\left(\frac{1}{x_1x_2\cdots x_{N-1}}\prod^{N-1}_{i< j}\frac{\prod^3_{b=1}\Gamma(x_{ij}^{\pm}f_b)}{\Gamma(x_{ij}^{\pm})}\prod^{N-1}_{i=1}
\frac{\prod^3_{b=1}\Gamma((x_1 \cdots x_i^2\cdots x_{N-1})^{\pm }f_b)}{\Gamma((x_1 \cdots x_i^2\cdots x_{N-1})^{\pm })}\right)\\
&=\prod^{N-1}_{i< j}\frac{\prod^3_{b=1}\Gamma((f_{a_i}f_{a_j}^{-1}p^{k_i-k_j}q^{l_i-l_j})^{\pm}f_b)}{\Gamma((f_{a_i}f_{a_j}^{-1}p^{k_i-k_j}q^{l_i-l_j})^{\pm})}\prod^{N-1}_{i=1}
\frac{\prod^3_{b=1}\Gamma((f_{a_i}p^{k_i}q^{l_i})^{\pm}f_b)}{\Gamma(p^{-k_i}q^{-l_i})\Gamma((f_{a_i}p^{k_i}q^{l_i})^{\pm })}\\
&\times \mathrm{Res}_{\sum p^{(n)}}\left(\prod^{N-1}_{i=1}\frac{(x_1 \cdots x_i^2\cdots x_{N-1})^{-1}f_{a_i})}{x_i}\right)\\
&=\prod^{N-1}_{i< j}\frac{\prod^3_{b=1}\Gamma((f_{a_i}f_{a_j}^{-1}p^{k_i-k_j}q^{l_i-l_j})^{\pm}f_b)}{\Gamma((f_{a_i}f_{a_j}^{-1}p^{k_i-k_j}q^{l_i-l_j})^{\pm})}\prod^{N-1}_{i=1}
\frac{\prod^3_{b=1}\Gamma((f_{a_i}p^{k_i}q^{l_i})^{\pm}f_b)}{\Gamma(p^{-k_i}q^{-l_i})\Gamma((f_{a_i}p^{k_i}q^{l_i})^{\pm })}\\
&\times \frac{1}{(p;p)^{N-1}_{\infty}(q;q)^{N-1}_{\infty}}\prod^{N-1}_{i=1}\frac{1}{C_i\prod^{k_i}_{m=1}\theta_q(p^{-m})\prod^{l_i}_{n=1}\theta_p(q^{-n})}
\end{split}
\end{align}
In the second line, second product, we kept the product over $b$ complete, even though $N-1$ of these factors are used in the residue.
The reason for this is to keep notation simple.
The price is that we have to add to the denominator a factor of $\Gamma(p^{-k_i}q^{-l_i})$ to cancel those superfluous factors in the numerator.
In addition, the $\theta$ functions in the last line originate from using the shift properties \eqref{eq:basic-shift-props-Gamma} for the $\Gamma$ functions contributing to the pole in order to evaluate the residue in terms of $q$-Pochhammer symbols.
This also results in $C_i$, which is given by:
\begin{equation}
C_i=\left(-p^{\frac{k_i+1}{2}}q^{\frac{l_i+1}{2}}\right)^{-k_il_i}.
\end{equation}

Summing the result over $k_i$ and $l_i$ gives us the final result for the class of poles \eqref{eq:pole-suN}.
Before stating the (form of) the final result, let us make the following remark.
As alluded to above, for specific values of $k_i$ and $l_i$, the residues may be vanishing.
This is caused by the $\Gamma$ functions in the denominator of the first factor of the residue.
To see this explicitly, let us first note that (see \eqref{eq:gamma-theta-id}):
\begin{align}\label{eq:gammas-as-thetas}
\begin{split}
&\frac{1}{\Gamma((f_{a_i}f_{a_j}^{-1}p^{k_i-k_j}q^{l_i-l_j})^{\pm})}=\theta_p(f_{a_i}f_{a_j}^{-1}p^{k_i-k_j}q^{l_i-l_j})\theta_q(f_{a_i}^{-1}f_{a_j}p^{k_j-k_i}q^{l_j-l_i})
\end{split}
\end{align}
Since the $a_i$ only takes three different values, for $N>4$ there are necessarily terms where the $f_{a_i}$ fugacities in the argument of the $\theta$ functions cancel.
If in addition for these values of $i$ and $j$ either $k_i=k_j$ or $l_i=l_j$, one of the $\theta$ functions on the right hand side has a zero.
Therefore, generically one has to require distinct $k_i$ and distinct $l_i$ for the residue to be non-vanishing.
In particular, this implies that the basic pole ($k_i,l_i=0$) necessarily has a vanishing residue for $N>4$.

We are now in a position to state the full residue sum for the class of poles \eqref{eq:pole-suN}:
\begin{align}\label{eq:suN-index-intermediate}
\begin{split}
I_N^{\prime}&=\frac{(\Gamma(f_1)\Gamma(f_2)\Gamma(f_3))^{N-1}}{N!}\sum^\prime_{(a_i)}\sum^\prime_{(k_i),(l_i)\geq (0)}\Bigg(\prod^{N-1}_{i< j}\frac{\prod^3_{b=1}\Gamma((f_{a_i}f_{a_j}^{-1}p^{k_i-k_j}q^{l_i-l_j})^{\pm}f_b)}{\Gamma((f_{a_i}f_{a_j}^{-1}p^{k_i-k_j}q^{l_i-l_j})^{\pm})}\\
&\times\prod^{N-1}_{i=1}
\frac{\prod^3_{b=1}\Gamma((f_{a_i}p^{k_i}q^{l_i})^{\pm}f_b)}{\Gamma(p^{-k_i}q^{-l_i})\Gamma((f_{a_i}p^{k_i}q^{l_i})^{\pm })}\frac{1}{C_i\prod^{k_i}_{m=1}\theta_q(p^{-m})\prod^{l_i}_{n=1}\theta_p(q^{-n})}\Bigg)
\end{split}
\end{align}
Before simplifying this expression, let us make some comments.
First of all, the sum over $(a_i)$ is similar to the sum over $a$ in the $SU(2)$ case and should be thought of as a sum over all possible $N-1$ tuples $(a_1,\ldots,a_{N-1})$ for $a_i=1,2,3$.
Similarly, by $(k_i)$, $(l_i)$ and $(0)$ we indicate the $N-1$ tuples $(k_1,\ldots,k_{N-1})$, $(l_1,\ldots,l_{N-1})$ and $(0,\ldots,0)$.
Secondly, the prime on the index indicates that we have only considered the poles \eqref{eq:pole-suN}.
Two other comments pertaining to the expression are:
\begin{enumerate}
\item As discussed around \eqref{eq:pole-suN-spec}, not all poles considered in the sum will fall inside all unit circles for generic values of the fugacities.
Poles which do not should not be included in the residue sum.
This necessary omission is indicated by the primes on both sums.
The explicit prescription would depend on where one evaluates the residue in fugacity space.
Finding this prescription seems a complicated problem, and we will not consider it in the present paper.
\item Similarly, we have not considered any of the poles where some of the $+$ signs are replaced with $-$ signs in the pole equations in the second line of \eqref{eq:lin-eqns-set}.
Some of these poles may still lie inside all unit circles in some regime of parameter space and therefore could be included into a full expression for the index.
\end{enumerate}
For our purposes, these two issues are not consequential for the following two reasons respectively:
\begin{enumerate}
\item To leading order in the Cardy limit of the index, we will show that there is a universal contribution (at large-$N$) for all poles (in a specific parameter regime).
Hence, to compute the Cardy limit one really only relies on the fact that the residue sum contains at least one non-vanishing residue.
\item Suppose there is a pole inside all unit circles after having replaced some of the $+$ signs with $-$ signs.
Then, one can obtain its residue from the residue of the pole with all $+$ signs upon using the transformation \eqref{eq:plus-to-min}.
Now, it is not difficult to see that the $\mathcal{O}(N^2)$ part of $I_N^\prime$, captured by the $\prod_{i<j}$ part of \eqref{eq:suN-index-intermediate}, is invariant under this transformation. 
Therefore, at large-$N$ the residues for this more general class of poles are indistinguishable from the residues of poles with only $+$ signs.
\end{enumerate}
With the above comments in mind, we proceed to simplify the expression.
We will make use again of the shift properties of the $\Gamma$ functions (see \eqref{eq:shift-props-gammas-app}).
Without further ado, we have:
\begin{align}\label{eq:simpl-gen-res-suN}
\begin{split}
&\prod^{N-1}_{i< j}\frac{\prod^3_{b=1}\Gamma((f_{a_i}f_{a_j}^{-1}p^{k_i-k_j}q^{l_i-l_j})^{\pm}f_b)}{\Gamma((f_{a_i}f_{a_j}^{-1}p^{k_i-k_j}q^{l_i-l_j})^{\pm})}\prod^{N-1}_{i=1}\Bigg(\frac{\prod^3_{b=1}\Gamma((f_{a_i}p^{k_i}q^{l_i})^{\pm}f_b)}{\Gamma(p^{-k_i}q^{-l_i})\Gamma((f_{a_i}p^{k_i}q^{l_i})^{\pm})}\\
&\qquad \qquad\qquad \qquad \qquad \qquad \qquad\qquad\qquad\times \frac{1}{C_i\prod^{k_i}_{m=1}\theta_q(p^{-m})\prod^{l_i}_{n=1}\theta_p(q^{-n})}\Bigg)\\
&=\prod^{N-1}_{i< j}\prod^3_{b=1}\Gamma((f_{a_i}f_{a_j}^{-1})^{\pm}f_b)\prod^{N-1}_{i=1}\frac{\prod^3_{b=1}\Gamma(f_{a_i}^{\pm}f_b)}{\Gamma(1)}\times  Z^{(k_i)}_{V}(\phi_{a_i},\sigma;\tau) Z^{(l_i)}_{V}(\phi_{a_i},\tau;\sigma).
\end{split}
\end{align}
The precise form of the vortex partition functions of the numerator $Z_{V}$ depends on the sign of $k_i-k_j$ and $l_i-l_j$.
For example, if both are positive or both negative for all $i<j$, then the vortex partition function is given by:\footnote{It is also possible to compute this partition function from the point of view of the vortex worldsheet theory (see \cite{Chen:2014rca,Poggi:2017kut} for examples of such a computation in $\mathcal{N}=1,2$ gauge theories). In particular, it should match the elliptic genus of a specific $(4,4)$ GLSM appearing for example in Section 5.1 of \cite{Gadde:2013dda}. Recently, the computation of the elliptic genus for a special example of the relevant GLSM appeared in \cite{Benini:2018hjy}.}
\begin{align}\label{eq:vortex-part-suN}
\begin{split}
 Z^{(k_i)}_{V}(\phi_{a_i},\sigma;\tau)&=\prod^{N-1}_{i< j}\frac{\prod^{k_i-k_j}_{m=1}\theta_q(f_{a_i}^{-1}f_{a_j}p^{-m})}{\prod^{k_i-k_j-1}_{m=1}\theta_q(f_{a_i}f_{a_j}^{-1}p^m)}\prod^{N-1}_{i=1}\frac{\prod^{k_i}_{m=1}\theta_q(f_{a_i}^{-1}p^{-m})}{\prod^{k_i-1}_{m=1}\theta_q(f_{a_i}p^m)}\\
&\times\prod^3_{b=1}\Bigg(\prod^{N-1}_{i< j}\frac{\prod^{k_i-k_j-1}_{m=0}\theta_q(f_{a_i}f_{a_j}^{-1}f_bp^m)}{\prod^{k_i-k_j}_{m=1}\theta_q(f_{a_i}^{-1}f_{a_j}f_bp^{-m})}
\prod^{N-1}_{i=1}\frac{\prod^{k_i-1}_{m=0}\theta_q(f_{a_i}f_bp^m)}{\prod^{k_i}_{m=1}\theta_q(f_{a_i}^{-1}f_bp^{-m})}\Bigg).
\end{split}
\end{align}
Here we used, as in the $SU(2)$ case, $f_1f_2f_3p^{-1}q^{-1}=1$.
Moreover, note that the products in the denominator of the first line start at $m=1$.
This is because the $m=0$ terms cancel the $\frac{1}{\Gamma((f_{a_i}f_{a_j}^{-1})^\pm)}$ and $\frac{1}{\Gamma(f_{a_i}^\pm)}$ terms, originating from the $\Gamma$ functions in the denominators of the first line of \eqref{eq:simpl-gen-res-suN}.
We find it more transparent to cancel these terms against each other directly, instead of defining $Z_V$ including the relevant $m=0$ terms and keeping the $\Gamma$ functions in the expression for the index.

For more general signs of $k_i-k_j$ and $l_i-l_j$, one has to make use also of the second and third line of \eqref{eq:shift-props-gammas-app}.
The net effect is a shuffling of $\theta$ functions between denominator and numerator.
Since this is not consequential for our purposes, we refrain from providing the precise formulas.
The precise form only depends on $(k_i)$ or $(l_i)$, and is summarized in that label on the vortex partition function.

Summing over all residues, we find a final expression for the index:\footnote{In this expression, $\Gamma(1)^{N-1}$ is only included to cancel the $\Gamma(1)^{N-1}$ coming from the last product in the first line.
These latter factors should not be included in the first place, since these represent precisely the factors that define the pole at which we evaluate the residue. The reason to include them is purely for notational convenience. This is related to the remarks in the paragraph below \eqref{eq:residue-basic-pole-suN}.}
\begin{align}\label{eq:suN-index-final}
\begin{split}
I_N^{\prime}&=\frac{(\Gamma(f_1)\Gamma(f_2)\Gamma(f_3))^{N-1}}{N!\Gamma(1)^{N-1}}\sum^\prime_{(a_i)}\prod^3_{b=1}\prod^{N-1}_{i< j}\Gamma((f_{a_i}f_{a_j}^{-1})^{\pm}f_b)\prod^{N-1}_{i=1}\Gamma((f_{a_i})^{\pm}f_b)\\
&\times \sum^\prime_{(k_i),(l_i)\geq (0)} Z^{(k_i)}_{V}(\phi_{a_i},\sigma;\tau) Z^{(l_i)}_{V}(\phi_{a_i},\tau;\sigma).
\end{split}
\end{align}
To repeat, there are two main provisos that should be kept in mind when reading this expression for the index. 
Firstly, the precise specification of the primed summation domain depends on where one is in parameter space.
This means that the final expression is not fully explicit.
Furthermore, we did not include any other classes of poles, corresponding to exchanges of $+$ sign equations with $-$ sign equations.
The first issue is problematic if one wants to study the index exactly.
However, we will see that for purposes of the Cardy limit it is not consequential at leading order.
The second issue can be easily resolved by inserting an additional sum over all transformations \eqref{eq:plus-to-min}.
However, also for these transformed residues, one would need to establish a summation domain, which will differ from the untransformed residue sum.
As mentioned in the main part of the section, the second issue disappears in the Cardy limit if combined with a large-$N$ limit, since the leading part of the residue at large-$N$ is universal for any combination of $+$ and $-$ sign equations.
Finally, as in the case of $SU(2)$, this expression is singular in the unrefined limit.
This is due to the fact that the integrand develops higher order poles, which require a separate analysis.
We perform this analysis in Appendix \ref{app:unrefined-limit-index}.

\section{Cardy limit of the index}\label{sec:cardy-limit-index}

In this section, we will study our final expression for the index \eqref{eq:suN-index-final} in the Cardy limit.
As commented above, even though the expression is not fully explicit, we will see that it suffices for our purposes.

\subsection{Cardy limit of \texorpdfstring{$\theta$}{theta} and \texorpdfstring{$\Gamma$}{Gamma} functions}\label{ssec:mod-prop-theta-gamma}

In order to study the Cardy limit of the index where $\tau,\sigma\to 0^{+i}$ keeping $\sigma/\tau\in \mathbb{H}\setminus \mathbb{R}$ fixed, we will make use of modular properties of $\theta$ and $\Gamma$ functions.
We will briefly review these properties here.
For a collection of relevant formulas, see also Appendix \ref{app:props-theta-gamma}.

\paragraph{$\theta$ function:}

\begin{figure}
	\centering
	\includegraphics[width=.5\textwidth]{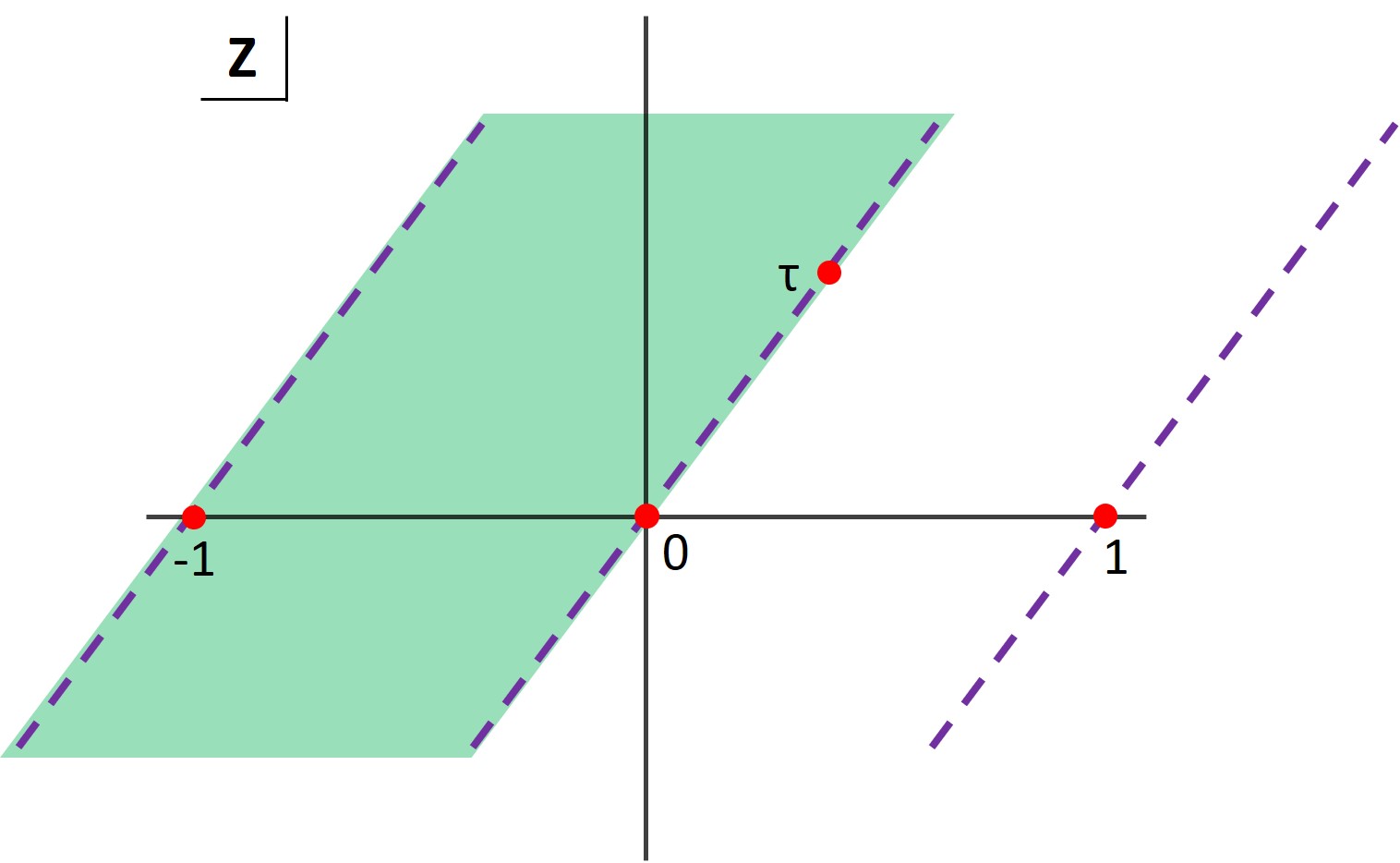}
	\caption{The complex $z$ plane. The shaded domain represents the domain $\mathrm{Im}\left(\frac{-1}{\tau}\right)>\mathrm{Im}\left(\frac{z}{\tau}\right)>0$.}
	\label{fig:z-regions}
\end{figure}

The $\theta$ function satisfies the following modular property:
\begin{equation}\label{eq:mod-prop-theta}
\theta(z;\tau)=e^{-i\pi B(z,\tau)}\theta\left(\frac{z}{\tau};-\frac{1}{\tau}\right),
\end{equation}
where $B(z,\tau)$ is defined in \eqref{eq:B-pol}.
Using the summation formula for the $\theta$ function \eqref{eq:theta-sum-form}, we have in the limit that $\tau\to 0$:
\begin{equation}
\theta\left(\frac{z}{\tau};-\frac{1}{\tau}\right)=\exp\left(-\sum^{\infty}_{l=1}\frac{1}{l}\frac{e^{2\pi i l\frac{z}{\tau}}+e^{2\pi il \frac{-z-1}{\tau}}}{1-e^{-2\pi i l\frac{1}{\tau}}}\right)\xrightarrow{\tau\to 0} 1
\end{equation}
if 
\begin{equation}
\mathrm{Im}\left(\frac{-1}{\tau}\right)>\mathrm{Im}\left(\frac{z}{\tau}\right)>0.
\end{equation}
This domain is illustrated in Figure \ref{fig:z-regions}.
In this domain, the modular property implies:
\begin{equation}\label{eq:cardy-limit-theta}
\lim_{\tau\to 0}\theta(z;\tau)=\lim_{\tau\to 0}e^{-i\pi B(z,\tau)},
\end{equation}
which we will call the Cardy limit of the $\theta$ function.

The domain where \eqref{eq:cardy-limit-theta} holds can be extended to $z\in \mathbb{C}\setminus \mathbb{Z}+\gamma$, where $\gamma$ is the line running through $0$ and $\tau$.
In the figure, such domains correspond to arbitrary horizontal integer shifts of the strip.
This extension is possible due to periodicity under $z\to z+1$ of the left hand side of \eqref{eq:mod-prop-theta}, which is reflected on the right side through:
\begin{align}\label{eq:quasi-period-B-theta}
\begin{split}
B(z+1,\tau)-B(z,\tau)&=2\frac{z+1}{\tau}-1,\\
\theta\left(\frac{z+1}{\tau};-\frac{1}{\tau}\right)&=-e^{\frac{2\pi i (z+1)}{\tau}}\theta\left(\frac{z}{\tau};-\frac{1}{\tau}\right).
\end{split}
\end{align}
The second line follows from the quasi-ellipticity of the $\theta$ function \eqref{eq:shift-prop-theta}.
One easily sees that the shift properties cancel in the product, thus reproducing the periodicity of the left hand side.
The extension of \eqref{eq:cardy-limit-theta} to $z\in \mathbb{C}\setminus \mathbb{Z}+\gamma$ can now be performed through repeated use of \eqref{eq:quasi-period-B-theta}.
The result is that for any $z\in \mathbb{C}\setminus \mathbb{Z}+\gamma$, we can write:
\begin{equation}\label{eq:cardy-limit-theta-gen-z}
\lim_{\tau\to 0}\theta(z;\tau)=\lim_{\tau\to 0}e^{-i\pi B([z]_\tau,\tau)},
\end{equation}
where the bracket is defined as:
\begin{equation}\label{eq:bracket-defn}
[z]_\tau\equiv z+n, \quad n\in \mathbb{Z} \quad \text{such that} \quad \mathrm{Im}\left(\frac{-1}{\tau}\right)>\mathrm{Im}\left(\frac{[z]_\tau}{\tau}\right)>0.
\end{equation}
In words, the bracket implements a horizontal shift on $z$ such that its image lies in the fundamental domain indicated in Figure \ref{fig:z-regions}.
It is easy to verify that the brackets satisfy the following relations:
\begin{equation}\label{eq:bracket-relations}
[z+m]_\tau=[z]_\tau, \quad m\in \mathbb{Z}, \quad [z+\tau]_\tau=[z]_\tau+\tau,\quad [-z]_\tau=-[z]_\tau-1.
\end{equation}

\paragraph{$\Gamma$ function:}

The elliptic $\Gamma$ function also satisfies a modular property \cite{Felder_2000}, as recently discussed as well in \cite{Gadde:2020bov}.
For $\mathrm{Im}(\tau),\mathrm{Im}(\sigma),\mathrm{Im}\left(\frac{\sigma}{\tau}\right)>0$, one has:\footnote{The domain can be extended to $\tau,\sigma,\frac{\tau}{\sigma}\in \mathbb{C}\setminus \mathbb{R}$, as we discuss in Appendix \ref{app:props-theta-gamma}. We will not require this extension in the following, and therefore stick with this domain where the product expressions for the elliptic $\Gamma$ functions appearing in the formula manifestly converge. In the same appendix, we also discuss the relation of this modular property to the one used in \cite{Gadde:2020bov}. In particular, we warn the reader that the $Q$ polynomial in \cite{Gadde:2020bov} is not the same as ours.}
\begin{equation}\label{eq:mod-prop-Gamma}
\Gamma(z;\tau,\sigma)=e^{i\frac{\pi}{3} Q(z,\tau,\sigma)}\frac{\Gamma\left(\frac{z}{\tau};\frac{\sigma}{\tau},-\frac{1}{\tau}\right)}{\Gamma\left(\frac{z-\tau}{\sigma};-\frac{\tau}{\sigma},-\frac{1}{\sigma}\right)}.
\end{equation}
Here, $Q(z,\tau,\sigma)$ is defined in Appendix \ref{app:props-theta-gamma}.

\begin{figure}
	\centering
	\includegraphics[width=.6\textwidth]{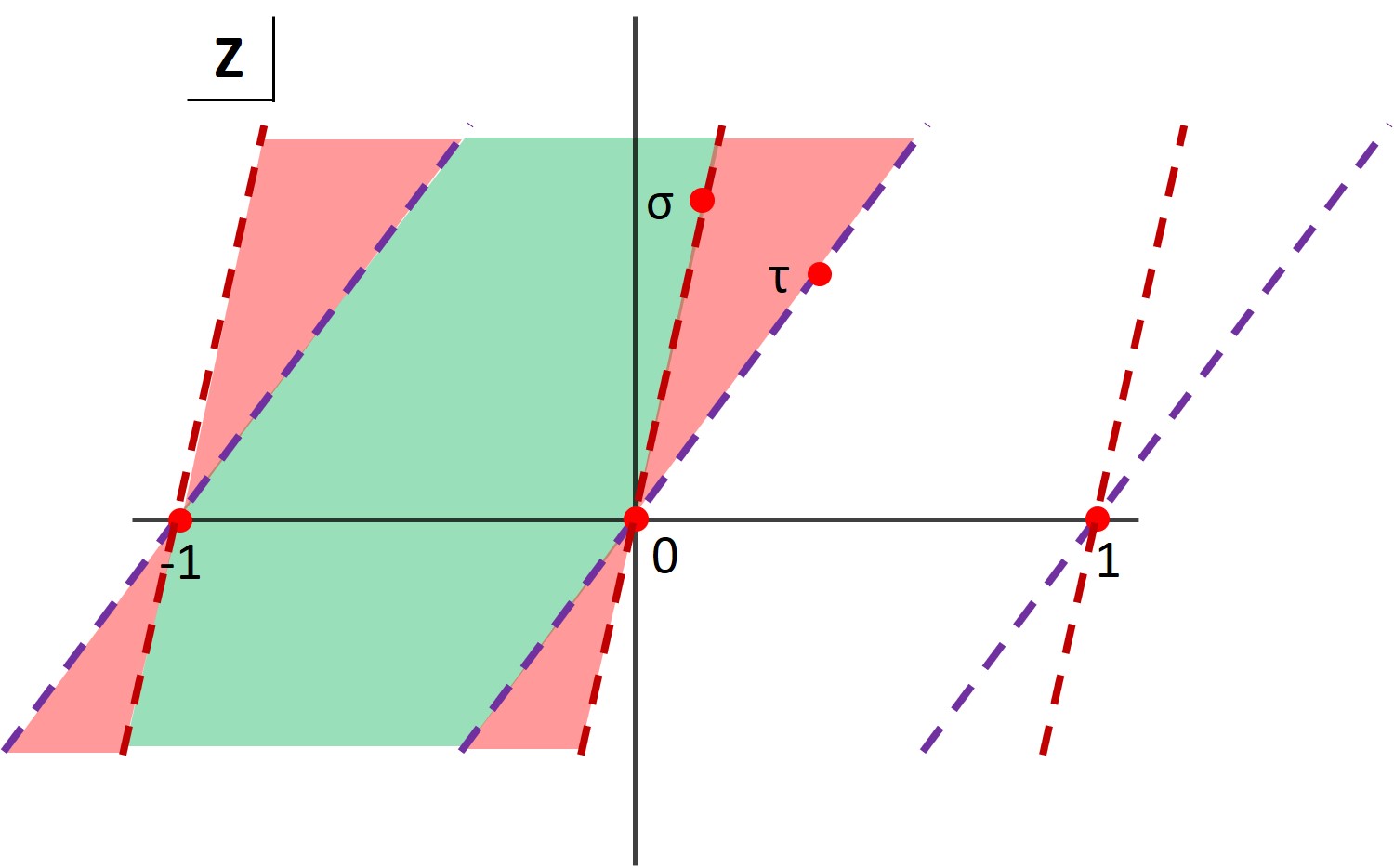}
	\caption{The complex $z$ plane. The shaded green domain represents the intersection of the domains $\mathrm{Im}\left(\frac{-1}{\tau}\right)>\mathrm{Im}\left(\frac{z}{\tau}\right)>0$ and $\mathrm{Im}\left(\frac{-1}{\sigma}\right)>\mathrm{Im}\left(\frac{z}{\sigma}\right)>0$. The shaded red domains belong to one domain but not both. The green domain is referred to as $D_0$ in the main text. Integer shifts to the left and right for $n>0$ are denoted by $D_n$ and $D_{-n}$ respectively.}
	\label{fig:z-regions-2}
\end{figure}

Following \cite{Gadde:2020bov}, consider the limit $\tau,\sigma\to 0$ with $\frac{\tau}{\sigma}\notin \mathbb{R}$ of the $\Gamma$ factor in the numerator on the right hand side of the modular property.
Using the summation formula \eqref{eq:defn-ell-gamma-sum}, we find:
\begin{align}\label{eq:cardy-lim-Gamma-factor}
\begin{split}
\Gamma\left(\frac{z}{\tau};\frac{\sigma}{\tau},-\frac{1}{\tau}\right)=&\exp\left(\sum^{\infty}_{l=1}\frac{1}{l}\frac{e^{2\pi il \frac{z}{\tau}}-e^{2\pi il \frac{-z+\sigma-1}{\tau}}}{(1-e^{2\pi il\frac{\sigma}{\tau}})(1-e^{-2\pi il\frac{1}{\tau}})}\right)\\
\xrightarrow{\tau,\sigma\to 0}&\exp\left(\sum^{\infty}_{l=1}\frac{1}{l}\frac{e^{2\pi il \frac{z}{\tau}}-e^{2\pi il \frac{-z-1}{\tau}}}{1-e^{2\pi il\frac{\sigma}{\tau}}}\right).
\end{split}
\end{align}
Similarly to the $\theta$ function, this factor becomes equal to $1$ when:
\begin{equation}\label{eq:tau-domain}
\mathrm{Im}\left(\frac{-1}{\tau}\right)>\mathrm{Im}\left(\frac{z}{\tau}\right)>0.
\end{equation}
The story is exactly the same for the $\Gamma$ function in the denominator if:
\begin{equation}
\mathrm{Im}\left(\frac{-1}{\sigma}\right)>\mathrm{Im}\left(\frac{z}{\sigma}\right)>0.
\end{equation}
The intersection of these two domains is shown in Figure \ref{fig:z-regions-2}.
For later convenience, we will refer to the green shaded diamond shaped domain as $D_{0}$.
We will refer to integer shifts of this domain by $D_{n}$, where $n\in \mathbb{Z}$ is defined in \eqref{eq:bracket-defn} and indicates a horizontal translation of $D_0$ by $-n$.

Thus, when the $z$ variable sits inside the diamond shaped green domain, the modular property reduces in the Cardy limit to:
\begin{equation}\label{eq:cardy-lim-gamma}
\lim_{\tau,\sigma\to 0}\Gamma(z;\tau,\sigma)=\lim_{\tau,\sigma\to 0}e^{i\frac{\pi}{3} Q(z,\tau,\sigma)}.
\end{equation}
Similarly to the case of the $\theta$ function, since $\Gamma(z;\tau,\sigma)$ is periodic under $z\to z+1$ this limit can be extended to $z\in D_n$ for any $n\in \mathbb{Z}$.
The periodicity of the left hand side of \eqref{eq:mod-prop-Gamma} is reflected on the right hand side through the following identities:
\begin{align}
\begin{split}
&Q(z+1,\tau,\sigma)-Q(z,\tau,\sigma)=-3\Phi(z+1,\tau,\sigma), \\
&\frac{\Gamma\left(\frac{z+1}{\tau};\frac{\sigma}{\tau},-\frac{1}{\tau}\right)}{\Gamma\left(\frac{z-\tau+1}{\sigma};-\frac{\tau}{\sigma},-\frac{1}{\sigma}\right)}=e^{i\pi\Phi(z+1,\tau,\sigma)}\frac{\Gamma\left(\frac{z}{\tau};\frac{\sigma}{\tau},-\frac{1}{\tau}\right)}{\Gamma\left(\frac{z-\tau}{\sigma};-\frac{\tau}{\sigma},-\frac{1}{\sigma}\right)}.
\end{split}
\end{align}
where we used the basic shift property \eqref{eq:basic-shift-gamma-app} of the elliptic $\Gamma$ function, the shift property \eqref{eq:shift-prop-theta} of the $\theta$ function, the extension \eqref{eq:theta-extend} of the $\theta$ function and finally the modular property \eqref{eq:mod-prop-theta-2} of the $\theta$ function.
In particular, the quadratic polynomial $\Phi(z,\tau,\sigma)$ in the chemical $z$ appears in the latter modular property and is defined in \eqref{eq:defn-Phi}.

Repeated use of these identities allows one to extend the limit \eqref{eq:cardy-lim-gamma} to $z\in D_n$ for any $n\in \mathbb{Z}$:
\begin{equation}\label{eq:cardy-limit-gamma-gen-z}
\lim_{\tau,\sigma\to 0}\Gamma(z;\tau,\sigma)=\lim_{\tau,\sigma\to 0}e^{i\frac{\pi}{3} Q([z],\tau,\sigma)},
\end{equation}
where the bracket is defined similarly as in \eqref{eq:bracket-defn}:
\begin{equation}
[z]\equiv z+n, \quad n\in \mathbb{Z} \quad \text{such that} \quad [z]\in D_0.
\end{equation}
However, in this case only the first and the third relation of \eqref{eq:bracket-relations} are satisfied:
\begin{equation}
[z+m]=[z],\quad m\in \mathbb{Z},\quad [-z]=-[z]-1.
\end{equation}
The second relation is not satisfied since enough translations by $\tau$ (or $\sigma$) will eventually bring a point inside the diamond into one of the red regions.
This will not be an issue for the remainder, since in the Cardy limit $\tau,\sigma\to 0$ and we can simply ignore their appearance in brackets to leading order.
Notice also that this bracket is only defined for $z\in D_n$.
This constrains the possible values of $z$ for which the Cardy limit results in \eqref{eq:cardy-limit-gamma-gen-z}, since the $D_n$ do not cover the full complex $z$-plane unless $\mathrm{arg}(\tau)=\mathrm{arg}(\sigma)$.

Let discuss the diamond domain $D_0$ in a little more detail.
In the following, we take $z=\phi_a$.
Depending on the sign of $\mathrm{Im}(\phi_a)$, the domain can be written as:
\begin{align}\label{eq:phi-domains}
\begin{split}
\mathrm{Im}(\phi_a)>0&:\quad\frac{\mathrm{Re}(\tau)}{\mathrm{Im}(\tau)}\mathrm{Im}(\phi_a)-1<\mathrm{Re}(\phi_a)<\frac{\mathrm{Re}(\sigma)}{\mathrm{Im}(\sigma)}\mathrm{Im}(\phi_a),\\
\mathrm{Im}(\phi_a)<0&:\quad\frac{\mathrm{Re}(\sigma)}{\mathrm{Im}(\sigma)}\mathrm{Im}(\phi_a)-1<\mathrm{Re}(\phi_a)<\frac{\mathrm{Re}(\tau)}{\mathrm{Im}(\tau)}\mathrm{Im}(\phi_a).
\end{split}
\end{align}
Notice that our choice $\mathrm{Im}\left(\frac{\sigma}{\tau}\right)$ implies that $\frac{\mathrm{Re}(\tau)}{\mathrm{Im}(\tau)}>\frac{\mathrm{Re}(\sigma)}{\mathrm{Im}(\sigma)}$ (see Figure \ref{fig:z-regions-2}).
It is then clear from these equations that the domain is maximal at $\mathrm{Im}(\phi_a)=0$ and shrinks linearly for both signs of $\mathrm{Im}(\phi_a)$, as is manifest from Figure \ref{fig:z-regions-2}.
In particular, the interval shrinks to zero size when:
\begin{align}
\begin{split}
\left|\mathrm{Im}(\phi_a)\right|=\frac{1}{\frac{\mathrm{Re}(\tau)}{\mathrm{Im}(\tau)}-\frac{\mathrm{Re}(\sigma)}{\mathrm{Im}(\sigma)}},
\end{split}
\end{align}
Therefore, if $\left|\mathrm{Im}(\phi_a)\right|$ is close to this value, $\phi_a$ will generically lie inside a red domain of Figure \ref{fig:z-regions-2}.
In this case, the divergence of the $\tau,\sigma\to 0$ limit cannot be isolated inside the $Q$ function as in \eqref{eq:cardy-limit-gamma-gen-z}.
Instead, one has to keep (one of) the elliptic $\Gamma$ functions on the right hand side of \eqref{eq:mod-prop-Gamma}.\footnote{We note here that there exist other modular properties for $\Gamma$ functions, for which a similar diamond domain exists that does not overlap with the current diamond. These modular properties will have different $Q$-polynomials, but can still be used to extend the applicability of formula like \eqref{eq:cardy-limit-gamma-gen-z}. We will explore such other modular properties in more detail in \cite{Jejjala:2021hlt}.}

To avoid restrictions on $\mathrm{Re}(\phi_a)$, we will consider the following regime in parameter space:
\begin{align}\label{eq:im-phi-constraint}
\begin{split}
\left|\mathrm{Im}(\phi_a)\right|\ll\frac{1}{\frac{\mathrm{Re}(\tau)}{\mathrm{Im}(\tau)}-\frac{\mathrm{Re}(\sigma)}{\mathrm{Im}(\sigma)}}.
\end{split}
\end{align}
This limit zooms into the part of the domain where the difference between the $\tau$ and $\sigma$ strip is very small. 
Effectively, in this regime we may treat the diamond as a strip and for generic values of $\phi_a$ in this regime: $\phi_a\in D_n$ for some $n$.
In this limit we can rewrite the domains in \eqref{eq:phi-domains} as follows:
\begin{align}
\begin{split}
\mathrm{Im}(\phi_a)>0&:\quad -1<\hat{\phi}^{+}_a<0,\\
\mathrm{Im}(\phi_a)<0&:\quad -1<\hat{\phi}^{-}_a<0,
\end{split}
\end{align}
where we defined scaled normal components $\hat{\phi}^+$  and $\hat{\phi}^-$ to the upper and lower right boundary of the diamond respectively:
\begin{align}\label{eq:phi-normal}
\begin{split}
\hat{\phi}^{+}_a&=\mathrm{Re}(\phi_a)-\frac{\mathrm{Re}(\sigma)}{\mathrm{Im}(\sigma)}\mathrm{Im}(\phi_a),\\
\hat{\phi}^{-}_a&=\mathrm{Re}(\phi_a)-\frac{\mathrm{Re}(\tau)}{\mathrm{Im}(\tau)}\mathrm{Im}(\phi_a).
\end{split}
\end{align}
Notice that for $\arg (\tau)$ and  $\arg (\sigma)$ close enough, \eqref{eq:im-phi-constraint} is not a strong constraint.
However, we do want the difference between $\arg (\tau)$ and  $\arg (\sigma)$ to be finite in order to use the modular property.


\subsection{Cardy limit}\label{ssec:cardy-limit-index}

We are now ready to compute the Cardy limit of our expression for the index \eqref{eq:suN-index-final}.
In the Cardy limit, the $\theta$ functions diverge as $\exp(\frac{1}{|\tau|})$ or $\exp(\frac{1}{|\sigma|})$, as can be seen from \eqref{eq:cardy-limit-theta-gen-z}.
On the other hand, the elliptic $\Gamma$ functions diverge as $\exp(\frac{1}{|\tau\sigma|})$, as shown in \eqref{eq:cardy-limit-gamma-gen-z}.
Therefore, to leading order we may ignore the $\theta$ functions and the difficulties associated to the sum over $(k_i)$ and $(l_i)$ commented upon at the end of Section \ref{ssec:suN-index}.
Instead, we only have to consider the part of the residues that is made up from $\Gamma$ functions:
\begin{equation}\label{eq:suN-index-univ-part}
I_N^{\prime}=\frac{(\Gamma(f_1)\Gamma(f_2)\Gamma(f_3))^{N-1}}{N!\Gamma(1)^{N-1}}\sum^\prime_{(a_i)}\prod^{N-1}_{i< j}\frac{\prod^3_{b=1}\Gamma((f_{a_i}f_{a_j}^{-1})^{\pm}f_b)}{\Gamma((f_{a_i}f_{a_j}^{-1})^{\pm})}\prod^{N-1}_{i=1}\frac{\prod^3_{b=1}\Gamma(f_{a_i}^{\pm}f_b)}{\Gamma(f_{a_i}^{\pm})}
\end{equation}
Here, we temporarily reinstated the $\Gamma$ functions in the denominators, which cancel against $\theta$ functions in the vortex partition functions (see the remark below \eqref{eq:vortex-part-suN}).
We warn the reader that this may cause confusion for two reasons.
Firstly, it falsely suggests the residues are all vanishing. 
This is because the $f_{a_i}$ can only take on three values, implying that for $N>4$:
\begin{equation}
\prod_{i<j}\frac{1}{\Gamma(f_{a_i}f_{a_j}^{-1})},
\end{equation}
necessarily has a zero. 
However, this is not a true zero since these factors are cancelled by the $m=0$ terms that we left out in the definition \eqref{eq:vortex-part-suN}. 
In addition, the rewriting is not necessary for taking the Cardy limit, again because these terms are cancelled by the $m=0$ $\theta$ functions.
In spite of this, we keep these $\Gamma$ functions for the moment, because they allow a nice derivation of the anomaly polynomial of the theory as we will now show.

We use the modular property \eqref{eq:mod-prop-Gamma} to replace all elliptic $\Gamma$ functions in \eqref{eq:suN-index-univ-part}.
In particular, every $\Gamma$ function contributes a certain $Q$ function to the overall prefactor.
Collecting all those $Q$ functions, we find:
\begin{align}\label{eq:Qtot-suN}
\begin{split}
Q_{\mathrm{tot}}(\phi_{a_{i}})&= (N-1)\left(-Q(0)+\sum^3_{c=1}Q(\phi_c)\right)\\
&+\sum_{\phi\in\lbrace \phi_{a_i},\phi_{a_{i}}-\phi_{a_{j}}\rbrace }\left[ \sum^3_{b=1}\Big(Q(\phi+\phi_b)+Q(-\phi+\phi_b)\Big)-Q(\phi)-Q(-\phi)\right],
\end{split}
\end{align}
where the summation runs over the set for $1\leq i<j\leq N-1$.
Interestingly, the full summand of the $\phi$ summation does not depend on $\phi$ if $\phi_3=\tau+\sigma-\phi_1-\phi_2-1$.
At this stage, it is not clear why we should choose $\phi_3$ as such.
Indeed, at the level of the index any integer $k$ could have been added: $\phi_3=\tau+\sigma-\phi_1-\phi_2+k$.
We will derive $k=-1$ later when considering the Cardy limit in a specific region of parameter space.
For now, let us just take $k=-1$ and in addition note that:
\begin{equation}
 -Q(0)+\sum^3_{c=1}Q(\phi_c)=-3\frac{\phi_1\phi_2\phi_3}{\tau\sigma}.
\end{equation}
Since the second line of \eqref{eq:Qtot-suN} does not depend on $\phi$ we may set $\phi=0$ to find: 
\begin{equation}\label{eq:anom-pol}
Q_{\mathrm{tot}}(\phi_{a_{i}})=-3(N^2-1)\frac{\phi_1\phi_2\phi_3}{\tau\sigma}.
\end{equation}
This object is almost identical to the supersymmetric Casimir energy \cite{Assel:2014paa,Assel:2015nca}, although the latter is defined for $k=0$ in $\phi_3$.
This apparently small distinction played a crucial role in the derivation of the AdS$_5$ black hole entropy of \cite{Cabo-Bizet:2018ehj} (see also \cite{Hosseini:2017mds}).\footnote{To compare the expression to the generalized supersymmetric Casimir energy of \cite{Cabo-Bizet:2018ehj}, note that their chemical potentials are related to ours times $2\pi i$.}
Furthermore, \eqref{eq:anom-pol} is very closely related to the anomaly polynomial of the $\mathcal{N}=4$ $SU(N)$ theory (see Appendix \ref{app:anomaly-pol}), as first observed in \cite{Bobev:2015kza}.
The fact that anomaly polynomials can be derived through the use of the modular property of elliptic $\Gamma$ functions was already known from several works, including \cite{Spiridonov:2012ww,Nieri:2015yia,Brunner:2016nyk} and discussed recently in detail in \cite{Gadde:2020bov}.

Notice that naively \eqref{eq:Qtot-suN} seems to depend on the choice of pole, i.e.\ a choice of $a_i$. 
However, the above shows that it is independent of this choice.
The interpretation of this fact, as for example appearing in \cite{Gadde:2020bov}, is that the residue sum can be thought of physically as a sum over Higgs branch vacua and the anomaly polynomial $Q_{\mathrm{tot}}$ should not depend on a specific vacuum.\footnote{Note that the $\mathcal{N}=4$ theory does not have a Higgs branch, so this terminology is inappropriate for the case at hand. Up to such semantics, however, we expect a similar argument to apply here.}
Finally, this object is also the so-called entropy function that upon a Legendre transformation leads to the correct black hole entropy \cite{Hosseini:2017mds} (see also \cite{Cabo-Bizet:2018ehj,Choi:2018hmj,Benini:2018ywd}).

However, as our analysis of the Cardy limit of the $\Gamma$ function has indicated, this is not yet the end of the story.
In order to evaluate the limit, we have to evaluate $Q_{\mathrm{tot}}$ on bracketed potentials (see \eqref{eq:cardy-limit-gamma-gen-z}).
We now only take the $Q$ polynomials of the $\Gamma$ functions appearing in \eqref{eq:suN-index-final}.
In other words, we ignore the $-Q(\phi)-Q(-\phi)$ part of \eqref{eq:Qtot-suN}, which originates from the $\Gamma$ functions in the denominator of \eqref{eq:suN-index-univ-part}.
Consistent with the identity \eqref{eq:gamma-theta-id}, this part is subleading in the Cardy limit:
\begin{equation}
Q([\phi])+Q([-\phi])=Q([\phi])+Q(-[\phi]-1)=\frac{(\tau+\sigma)(6[\phi]^2+6[\phi]+1+\tau\sigma)}{2\tau\sigma},
\end{equation}
where we made use of the bracket relations \eqref{eq:bracket-relations}.

Thus, the total polynomial to consider is now given by:
\begin{align}\label{eq:Qtot-prime-suN}
\begin{split}
Q^{\prime}_{\mathrm{tot}}(\phi_{a_{i}})&= (N-1)\left(-Q(0)+\sum^3_{b=1}Q([\phi_b])\right)\\
&+\sum_{\phi\in\lbrace \phi_{a_i},\phi_{a_{i}}-\phi_{a_{j}}\rbrace} \sum^3_{b=1}Q([\phi+\phi_b])+Q([-\phi+\phi_b]),
\end{split}
\end{align}
Before turning to an analysis of this object as a function of the specific residue, we will make some comments:
\begin{enumerate}
\item At large-$N$, one may just consider the summation over $\phi_{a_{i}}-\phi_{a_{j}}$ since only this part contributes $\mathcal{O}(N^2)$ terms.
\item The brackets will reintroduce a dependence on the summation variable $\phi$, in contrast with the analysis of \eqref{eq:Qtot-suN}.
In principle, this implies that different residues contribute differently in the Cardy limit.
Notice that this presents an opposing point of view to the analysis in \cite{Gadde:2020bov}, where it is asserted that also in the Cardy limit, the residues/vacua contribute universally.
\item We will show below that the latter point can be partially resolved by taking $N$ large and restricting the flavor fugacities appropriately.
\end{enumerate}

\subsubsection{Equal \texorpdfstring{$a_i$}{ai}}

Let us start with the analysis of a residue for which $a_{i}=a_{j}$ for all $i,j$.
Focusing on the part that scales with $N^2$,\footnote{We will comment on subleading corrections in $N$ at the end of this section.} we find:
\begin{equation}
Q^{\prime}_{\mathrm{tot}}(\phi_{a_{i}})\approx N^2 \sum^3_{b=1}Q([\phi_b])=N^2\Big(Q([\phi_1])+Q([\phi_2])+Q([\tau+\sigma-\phi_1-\phi_2])\Big).
\end{equation}
To evaluate the bracket appearing in the last $Q$ function, we first note that we may ignore $\tau$ and $\sigma$ to leading order in the Cardy limit.
Now, there are two possibilities depending on whether $[\phi_1]+[\phi_2]\in D_{0,1}$ (see Figure \ref{fig:z-regions-2}):
\begin{equation}
[\phi_1+\phi_2]=\begin{cases}
[\phi_1]+[\phi_2]  &\text{if}\quad [\phi_1]+[\phi_2]\in D_0\\
[\phi_1]+[\phi_2]+1 &\text{if}\quad [\phi_1]+[\phi_2]\in D_1.
\end{cases}
\end{equation}
If instead $[\phi_1]+[\phi_2]\notin D_{0,1}$ we cannot proceed, as we explained at the end of Section \ref{ssec:mod-prop-theta-gamma}.
The two choices lead respectively to:
\begin{equation}
Q^{\prime}_{\mathrm{tot}}(\phi_{a_{i}})\approx \begin{cases}
 3N^2 \frac{[\phi_1][\phi_2]([\phi_1]+[\phi_2]+1)}{\tau\sigma}+\mathcal{O}(\tau^{-1})+\mathcal{O}(\sigma^{-1})  &\text{if}\quad [\phi_1]+[\phi_2]\in D_0\\
 3N^2 \frac{[\phi_1]^\prime [\phi_2]^\prime([\phi_1]^\prime+[\phi_2]^\prime-1)}{\tau\sigma}+\mathcal{O}(\tau^{-1})+\mathcal{O}(\sigma^{-1}) &\text{if}\quad [\phi_1]+[\phi_2]\in D_1,
\end{cases} 
\end{equation}
where we defined:
\begin{equation}
[\phi_a]^\prime=[\phi_a]+1.
\end{equation}
We can make the expression look more symmetric by using $[\phi_3]=[-\phi_1-\phi_2]$.
In this case, the expression becomes:
\begin{equation}\label{eq:entropy-function-ai=aj}
Q^{\prime}_{\mathrm{tot}}(\phi_{a_{i}})\approx \begin{cases}
 -3N^2 \frac{[\phi_1][\phi_2][\phi_3]}{\tau\sigma}+\mathcal{O}(\tau^{-1})+\mathcal{O}(\sigma^{-1})  &\text{if}\quad [\phi_1]+[\phi_2]\in D_0\\
- 3N^2 \frac{[\phi_1]^\prime [\phi_2]^\prime[\phi_3]^\prime}{\tau\sigma}+\mathcal{O}(\tau^{-1})+\mathcal{O}(\sigma^{-1}) &\text{if}\quad [\phi_1]+[\phi_2]\in D_1,
\end{cases} 
\end{equation}
This is the expected answer for the entropy function \eqref{eq:anom-pol},\footnote{Notice that there is a subtle difference between the Cardy limit of the index and the entropy function. Indeed, the Cardy limit of the index is periodic in the $\phi_a$ whereas the entropy function is not.
This was mentioned and resolved very recently in a talk by Ofer Aharony \cite{AharonyTalk}. In the following, we will have this resolution in mind when comparing our results to the entropy functions.\label{fn:ofer}} where the two possibilities correspond to the twin saddles discussed for the first time in \cite{Cabo-Bizet:2018ehj,Choi:2018hmj,Benini:2018ywd}.
In particular, a Legendre transform of both expressions gives rise the correct black hole entropy \cite{Hosseini:2017mds} (see also \cite{Cabo-Bizet:2018ehj,Choi:2018hmj,Benini:2018ywd}).
Even though this is encouraging, we still have to analyse more general residues before we can make definite statements about the Cardy limit of the full index.

\subsubsection{Unequal \texorpdfstring{$a_i$}{ai}}

To understand if there is a universal contribution from all residues to the index in the Cardy limit, or if some residues are subleading with respect to the residue studied above, we will now analyse the most general residue.
Apart from the terms in the sum over $\phi_{a_i}-\phi_{a_j}$ where $a_i=a_j$, there are three other possible terms contributing at leading order in $N$, corresponding to the pairs $(a_i,a_j)=(1,2),(1,3),(2,3)$.
The sums over $b=1,2,3$ in \eqref{eq:Qtot-prime-suN} work out for each pair respectively as:
\begin{align}\label{eq:Qs-phia-phib}
\begin{split}
Q_{12}&= Q([2\phi_1-\phi_2])+Q([\phi_2])+Q([\phi_1])+Q([2\phi_2-\phi_1])+Q([-2\phi_2])\\
&+Q([-2\phi_1]),\\
Q_{13}&= Q([3\phi_1+\phi_2])+Q([-\phi_1-\phi_2])+Q([2\phi_1+2\phi_2])+Q([-2\phi_1])+Q([\phi_1])\\
&+Q([-3\phi_1-2\phi_2]),\\
Q_{23}&= Q([\phi_1+3\phi_2])+Q([-\phi_1-\phi_2])+Q([2\phi_1+2\phi_2])+Q([-2\phi_2])+Q([\phi_2])\\
&+Q([-2\phi_1-3\phi_2]).
\end{split}
\end{align}
To proceed, we first have to understand how to evaluate the more general brackets appearing in the arguments of the $Q$ functions.
For a bracket $[a\phi_1+b\phi_2]$ with $a,b\geq 0$, the following cases should be considered:
\begin{equation}\label{eq:bracket-ab}
[a\phi_1+b\phi_2]=\begin{cases}
a[\phi_1]+b[\phi_2] & \mathrm{if} \quad a[\phi_1]+b[\phi_2]\in D_0,\\
a[\phi_1]+b[\phi_2]+1 & \mathrm{if} \quad a[\phi_1]+b[\phi_2]\in D_1,\\
\qquad \;\;\vdots \\
a[\phi_1]+b[\phi_2]+a+b-1 & \mathrm{if} \quad a[\phi_1]+b[\phi_2]\in D_{a+b-1}.
\end{cases}
\end{equation}
In addition, the brackets $[2\phi_{1,2}-\phi_{2,1}]$ can be evaluated as follows:
\begin{equation}\label{eq:extra-brackets}
[2\phi_{1,2}-\phi_{2,1}]=\begin{cases}
2[\phi_{1,2}]-[\phi_{2,1}]-1& \mathrm{if} \quad 2[\phi_{1,2}]-[\phi_{2,1}]\in D_{-1},\\
2[\phi_{1,2}]-[\phi_{2,1}]&\mathrm{if} \quad 2[\phi_{1,2}]-[\phi_{2,1}] \in D_{0},\\
2[\phi_{1,2}]-[\phi_{2,1}]+1& \mathrm{if} \quad 2[\phi_{1,2}]-[\phi_{2,1}]+1 \in D_{1}.
\end{cases}
\end{equation}
If any of the brackets do not fall within any diamond, we cannot proceed.
As explained in Section \ref{ssec:mod-prop-theta-gamma}, we can avoid this issue by working in the limit \eqref{eq:im-phi-constraint}.
In the following, we will always work in this limit.

\begin{figure}
	\centering
	\includegraphics[width=.5\textwidth]{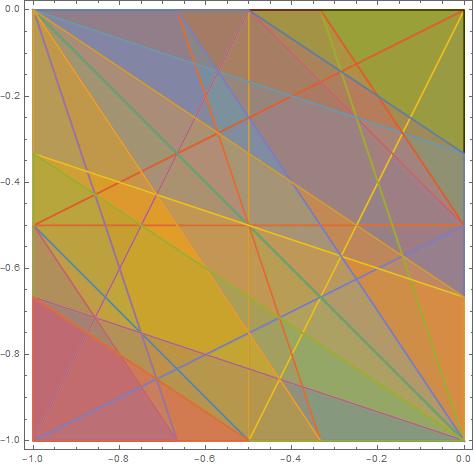}
	\caption{The various domains in $[\hat{\phi}^{\pm}_{1,2}]$ space for which the added constants to the brackets in \eqref{eq:Qs-phia-phib} take specific values. For example, all added integers vanish in the domain enclosed by the red and yellow line in the upper right corner. As one crosses domain borders, an integer in one of the ten brackets appearing in \eqref{eq:Qs-phia-phib} changes.}
	\label{fig:Q-phia-phib-domains}
\end{figure}

We will now compute the value of the added constant integer for all brackets as a function of $-1<[\hat{\phi}^{\pm}_{1,2}]<0$, where $\hat{\phi}^{\pm}_{1,2}$ were defined in \eqref{eq:phi-normal}.
As functions of $\hat{\phi}^{\pm}_{1,2}$ each separate bracket $[a\phi_1+b\phi_2]$ divides the square $-1<[\hat{\phi}^{\pm}_{1,2}]<0$ up into $|a|+|b|$ parallel strips. 
Superimposing the strips for all ten distinct brackets appearing in \eqref{eq:Qs-phia-phib} yields the various domains where the brackets take on a distinct set of values, as illustrated in Figure \ref{fig:Q-phia-phib-domains}.

To find out whether there exists a universal residue in some region of $\phi_a$ space, i.e.\ a point/region where $Q_{12}=Q_{13}=Q_{23}$, it is natural to start with the unrefined point(s) $[\phi]\equiv[\phi_1]=[\phi_2]=[\phi_3]$.
This equation has two solutions:
\begin{align}\label{eq:unref-points}
\begin{split}
[\phi]=-\tfrac{1}{3} \quad &\text{if}\quad [\phi_1]+[\phi_2]\in D_0\\
[\phi]=-\tfrac{2}{3}\quad &\text{if}\quad [\phi_1]+[\phi_2]\in D_1.
\end{split} 
\end{align}
Notice that this corresponds to the usual unrefined point $\phi=\tfrac{1}{3}\left(\tau+\sigma \pm 1\right)$ (up to a translation to $D_0$) in the Cardy limit.
Also, note that these points obey the requirement \eqref{eq:im-phi-constraint}.
At these points $a_i\neq a_j$ is irrelevant, and for the same reasons as in the case when $a_i=a_j$ for all $i,j$ we obtain the entropy functions \eqref{eq:entropy-function-ai=aj} (see footnote \ref{fn:ofer}), now evaluated at the two unrefined points in \eqref{eq:unref-points}, respectively.
We should note here that naively we are not allowed to evaluate \eqref{eq:Qtot-prime-suN} at the unrefined points.
This is due to the fact that at the unrefined points our expression for the index is singular because the integrand develops higher poles in this limit.
However, at large-$N$, i.e.\ when restricting to the $\phi\in\lbrace \phi_{a_i}-\phi_{a_j}\rbrace$ terms in \eqref{eq:Qtot-prime-suN}, we can safely take the unrefined limit of $Q^{\prime}_{\mathrm{tot}}$.
This is because this part of $Q^{\prime}_{\mathrm{tot}}$ does not change when taking the higher order poles into account, as we show explicitly in Appendix \ref{app:unrefined-limit-index}.
We will return to finite $N$ at the end of this section, where we will see how the $\mathcal{O}(N)$ contributions to $Q^{\prime}_{\mathrm{tot}}$ prohibit its evaluation on the unrefined point and we instead have to resort to the expression for $Q^{\prime}_{\mathrm{tot}}$ in Appendix \ref{app:unrefined-limit-index}.
Concluding, at least in the unrefined limits and at large-$N$, our expression for the index in the Cardy limit reproduces the expected entropy functions.

\begin{figure}
	\centering
	\includegraphics[width=.4\textwidth]{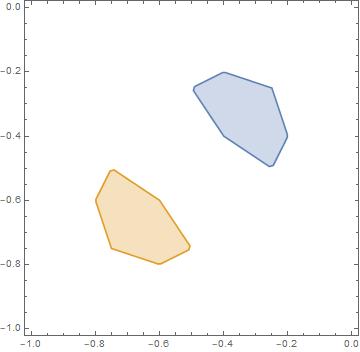}
	\caption{The blue and yellow domains correspond to the domains in Figure \ref{fig:Q-phia-phib-domains} that contain the unrefined points $[\phi]=-\tfrac{1}{3}$ and $[\phi]=-\tfrac{2}{3}$ respectively.}
	\label{fig:large-n-domain}
\end{figure}

Let us now try to move away from the unrefined point.
Before considering other domains, let us first just consider the domains that contain the unrefined points.
These domains are given by the blue and yellow domain in Figure \ref{fig:large-n-domain}.

In the blue domain, the (non-trivial) brackets evaluate as follows:
\begin{align}\label{eq:large-n-brackets}
\begin{split}
(1,2): &\;[2\phi_1-\phi_2]=2[\phi_1]-[\phi_2],\quad [2\phi_2-\phi_1]=2[\phi_2]-[\phi_1],\\
& [-2\phi_2]=-2[\phi_2]-1,[-2\phi_1]=-2[\phi_1]-1,\\
(1,3): &\;[3\phi_1+\phi_2]=3[\phi_1]+[\phi_2]+1,\quad [-\phi_1-\phi_2]=-[\phi_1]-[\phi_2]-1,\\
&[2\phi_1+2\phi_2]=2[\phi_1]+2[\phi_2]+1,\quad [-2\phi_1]=-2[\phi_1]-1,\\
&[-3\phi_1-2\phi_2]=-3[\phi_1]-2[\phi_2]-2,\\
(2,3):&\; [\phi_1+3\phi_2]=[\phi_1]+3[\phi_2]+1,\quad [-\phi_1-\phi_2]=-[\phi_1]-[\phi_2]-1,\\
&[2\phi_1+2\phi_2]=2[\phi_1]+2[\phi_2]+1,\quad\;[-2\phi_2]=-2[\phi_2]-1,\\
&[-2\phi_1-3\phi_2]=-2[\phi_1]-3[\phi_2]-2.
\end{split}
\end{align}
Evaluating the $Q_{ab}$ polynomials on these brackets, we find:
\begin{align}
\begin{split}
 Q_{12}&=\frac{6[\phi_1][\phi_2]([\phi_1]+[\phi_2]+1)}{\tau\sigma}-\frac{3([\phi_1]-[\phi_2])^2}{\tau\sigma}+\mathcal{O}(\tau^{-1})+\mathcal{O}(\sigma^{-1}),\\
 Q_{13}&=\frac{6[\phi_1][\phi_2]([\phi_1]+[\phi_2]+1)}{\tau\sigma}-\frac{3(1+2[\phi_1]+[\phi_2])^2}{\tau\sigma}+\mathcal{O}(\tau^{-1})+\mathcal{O}(\sigma^{-1}),\\
 Q_{23}&=\frac{6[\phi_1][\phi_2]([\phi_1]+[\phi_2]+1)}{\tau\sigma}-\frac{3(1+[\phi_1]+2[\phi_2])^2}{\tau\sigma}+\mathcal{O}(\tau^{-1})+\mathcal{O}(\sigma^{-1}).
\end{split}
\end{align}
Again, we can write this more symmetrically in terms of $[\phi_3]=-[\phi_1]-[\phi_2]-1$ as follows:
\begin{align}
\begin{split}
 Q_{12}&=-\frac{6[\phi_1][\phi_2][\phi_3]}{\tau\sigma}-\frac{3([\phi_1]-[\phi_2])^2}{\tau\sigma}+\mathcal{O}(\tau^{-1})+\mathcal{O}(\sigma^{-1}),\\
 Q_{13}&=-\frac{6[\phi_1][\phi_2][\phi_3]}{\tau\sigma}-\frac{3([\phi_1]-[\phi_3])^2}{\tau\sigma}+\mathcal{O}(\tau^{-1})+\mathcal{O}(\sigma^{-1}),\\
 Q_{23}&=-\frac{6[\phi_1][\phi_2][\phi_3]}{\tau\sigma}-\frac{3([\phi_2]-[\phi_3])^2}{\tau\sigma}+\mathcal{O}(\tau^{-1})+\mathcal{O}(\sigma^{-1}).
\end{split}
\end{align}
This result shows that as we move away from the unrefined point, as long as we stay in a small enough region around the unrefined point $[\phi_1]=[\phi_2]=-\tfrac{1}{3}$, we still obtain a universal residue up to small corrections of order $(\phi_{a}-\phi_{b})^2$.
That is, for this region in parameter space and at large-$N$ we have:
\begin{align}\label{eq:Qprime-gen-res}
\begin{split}
Q^{\prime}_{\mathrm{tot}}(\phi_{a_{i}})&\approx \sum_{\phi\in\lbrace \phi_{a_{i}}-\phi_{a_{j}}\rbrace} \sum^3_{b=1}Q([\phi+\phi_b])+Q([-\phi+\phi_b])\\
&=3N^2 \frac{[\phi_1][\phi_2][\phi_3]}{\tau\sigma}+\mathcal{O}(\tau^{-1})+\mathcal{O}(\sigma^{-1})+\mathcal{O}(([\phi_{a}]-[\phi_{b}])^2),
\end{split}
\end{align}
where the pair $(a,b)$ takes the values $(1,2)$, $(1,3)$ and $(2,3)$.
We stress that we did not have to impose $\phi_3=-\phi_1-\phi_2-1$.
Instead, the integer $-1$ emerges from a careful limit of the modular property and the associated bracketed potentials.
This is very similar to the emergence of this integer in the analysis of \cite{Benini:2018ywd}.

We can repeat the above analysis for the yellow domain in Figure \ref{fig:large-n-domain}.
Now, the unrefined point $[\phi]=-\tfrac{2}{3}$ is enclosed in the domain.
For this domain, the non-trivial brackets take the following form:
\begin{align}\label{eq:large-n-brackets-2}
\begin{split}
(1,2): &\;[2\phi_1-\phi_2]=2[\phi_1]-[\phi_2],\quad [2\phi_2-\phi_1]=2[\phi_2]-[\phi_1],\\
& [-2\phi_2]=-2[\phi_2]-2,[-2\phi_1]=-2[\phi_1]-2,\\
(1,3): &\;[3\phi_1+\phi_2]=3[\phi_1]+[\phi_2]+2,\quad [-\phi_1-\phi_2]=-[\phi_1]-[\phi_2]-2,\\
&[2\phi_1+2\phi_2]=2[\phi_1]+2[\phi_2]+2,\quad [-2\phi_1]=-2[\phi_1]-2,\\
&[-3\phi_1-2\phi_2]=-3[\phi_1]-2[\phi_2]-4,\\
(2,3):&\; [\phi_1+3\phi_2]=[\phi_1]+3[\phi_2]+2,\quad [-\phi_1-\phi_2]=-[\phi_1]-[\phi_2]-2,\\
&[2\phi_1+2\phi_2]=2[\phi_1]+2[\phi_2]+2,\quad\;[-2\phi_2]=-2[\phi_2]-2,\\
&[-2\phi_1-3\phi_2]=-2[\phi_1]-3[\phi_2]-4.
\end{split}
\end{align}
Plugging these brackets into \eqref{eq:Qtot-prime-suN}, we find:
\begin{align}\label{eq:Qs-phia-phib-2}
\begin{split}
Q^\prime_{12}&=\frac{6[\phi_1]^\prime[\phi_2]^\prime([\phi_1]^\prime+[\phi_2]^\prime-1)}{\tau\sigma}+\frac{3([\phi_1]^\prime-[\phi_2]^\prime)^2}{\tau\sigma}+\mathcal{O}(\tau^{-1})+\mathcal{O}(\sigma^{-1}),\\
 Q^\prime_{13}&=\frac{6[\phi_1]^\prime[\phi_2]^\prime([\phi_1]^\prime+[\phi_2]^\prime-1)}{\tau\sigma}+\frac{3(-1+2[\phi_1]^\prime+[\phi_2]^\prime)^2}{\tau\sigma}+\mathcal{O}(\tau^{-1})+\mathcal{O}(\sigma^{-1}),\\
 Q^\prime_{23}&=\frac{6[\phi_1]^\prime[\phi_2]^\prime([\phi_1]^\prime+[\phi_2]^\prime-1)}{\tau\sigma}+\frac{3(-1+[\phi_1]^\prime+2[\phi_2]^\prime)^2}{\tau\sigma}+\mathcal{O}(\tau^{-1})+\mathcal{O}(\sigma^{-1}),
\end{split}
\end{align}
where we remind the reader that:
\begin{equation}
[\phi_a]^\prime=[\phi_a]+1.
\end{equation}
In terms of $[\phi_3]'=-[\phi_1]'-[\phi_2]'+1$, this becomes:
\begin{align}\label{eq:Qs-phia-phib-phi3}
\begin{split}
Q^\prime_{12}&=\frac{6[\phi_1]^\prime[\phi_2]^\prime[\phi_3]^\prime}{\tau\sigma}+\frac{3([\phi_1]'-[\phi_2]')^2}{\tau\sigma}+\mathcal{O}(\tau^{-1})+\mathcal{O}(\sigma^{-1}),\\
 Q^\prime_{13}&=\frac{6[\phi_1]^\prime[\phi_2]^\prime[\phi_3]^\prime}{\tau\sigma}+\frac{3([\phi_1]'-[\phi_3]')^2}{\tau\sigma}+\mathcal{O}(\tau^{-1})+\mathcal{O}(\sigma^{-1}),\\
 Q^\prime_{23}&=\frac{6[\phi_1]^\prime[\phi_2]^\prime[\phi_3]^\prime}{\tau\sigma}+\frac{3([\phi_2]'-[\phi_3]')^2}{\tau\sigma}+\mathcal{O}(\tau^{-1})+\mathcal{O}(\sigma^{-1}),
\end{split}
\end{align}
Similarly to the blue region, all remainder terms are small close to the unrefined point $[\phi_1]=[\phi_2]=-\tfrac{2}{3}$.
Therefore, close enough to the unrefined point we are also able in this case to conclude that each residue contributes a universal $Q$ function in the Cardy limit and at large-$N$ up to small corrections:
\begin{align}\label{eq:Qprime-gen-res-twin}
\begin{split}
Q^{\prime}_{\mathrm{tot}}(\phi_{a_{i}})&\approx \sum_{\phi\in\lbrace \phi_{a_{i}}-\phi_{a_{j}}\rbrace} \sum^3_{b=1}Q([\phi+\phi_b])+Q([-\phi+\phi_b])\\
&=3N^2 \frac{[\phi_1]^\prime[\phi_2]^\prime[\phi_3]^\prime}{\tau\sigma}+\mathcal{O}(\tau^{-1})+\mathcal{O}(\sigma^{-1})+\mathcal{O}(([\phi_{a}]'-[\phi_{b}]')^2).
\end{split}
\end{align}
This function coincides with the entropy function for the twin saddle (again, see footnote \ref{fn:ofer}).

\subsubsection{Other domains}

Apart from the blue and yellow domains of Figure \ref{fig:large-n-domain}, there are many more domains in Figure \ref{fig:Q-phia-phib-domains}. 
In each of these domains, the set of brackets \eqref{eq:large-n-brackets} takes on a different value.
Similarly as for the blue and yellow regions, we can ask for each of these regions whether they contain a point where all the $Q_{ab}$ are equal to each other.
This would be indicative of the existence of a universal residue in such a region.
We find that there are two possible scenarios.
In the first scenario, which occurs for most domains, the point at which the $Q_{ab}$ associated to a certain region are equal falls outside that region.
Therefore, it seems that we cannot associate a universal residue to these regions.

Another scenario occurs for those domains whose boundary touches or overlaps with the lines $[\phi_1]=0$, $[\phi_2]=0$, $[\phi_1]^\prime=0$, $[\phi_2]^\prime=0$ or $[\phi_1]+[\phi_2]+1=0$.
In these cases, all $Q_{ab}$ are equal and vanishing along the intersection.
Interestingly, the entropy function for $\phi_1+\phi_2\in D_0$ vanishes along $\phi_1=0$, $\phi_2=0$ and $\phi_1+\phi_2+1=0$ whereas the entropy function for $\phi_1+\phi_2\in D_1$ vanishes $\phi_1+1=0$, $\phi_2+1=0$ and $\phi_1+\phi_2+1=0$.
Therefore, for these special regions we find that our result is again consistent with the known entropy functions (see footnote \ref{fn:ofer}).
However, we find that moving away from the vanishing locus generates remainder terms for at least one of the $Q_{ab}$ that are of the same order as the entropy function.
Therefore, we cannot keep the remainder terms for all three $Q_{ab}$ small while moving away from the vanishing locus, in contrast to the case of the blue and yellow region in Figure \ref{fig:large-n-domain}.

Concluding, we note that our expression for the index singles out the blue and yellow regions: only in these regions, close enough to the unrefined points, do we find a universal residue.
Moreover, this universal residue is consistent with the results in the literature.
An important difference with the Bethe Ansatz analysis is that in their case in the relevant parts of parameter space only a single residue dominates \cite{Benini:2018ywd}.
 
For the other domains we remain inconclusive because there does not seem to exist a universal residue.
This could mean that, instead of being able to extract such a universal piece, one would have to sum the various residues to find the Cardy limit.
This seems a very complicated task.
Or one may be able to argue that some residue provides the dominant contribution to the residue sum in the respective domain, similar to \cite{Benini:2018ywd}.
If a dominant residue would correspond to a residue with all $a_i=a_j$, we would get the expected entropy functions for either the right upper triangular region or lower left triangular region in Figure \ref{fig:Q-phia-phib-domains}, consistent with the literature.
However, we have so far not been able to find a convincing argument for this scenario.
Finally, let us note that whether a given residue will contribute to the residue sum depends in particular on the values of $|f_a|$, as we have seen in Section \ref{ssec:suN-index}. 
To fully understand the question of the existence of a universal residue in different regions of parameter space, one should first know which residues contribute in the first place. 
We do not expect this to fully resolve the issue, but further analysis of this point is required.

\subsection{A brief look at finite \texorpdfstring{$N$}{N}}

\begin{figure}
	\centering
	\includegraphics[width=.4\textwidth]{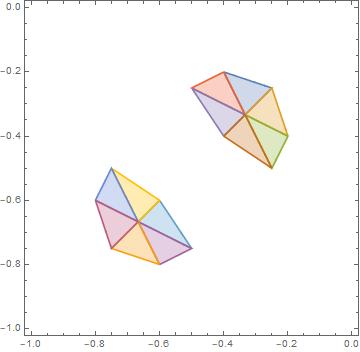}
	\caption{The additional brackets $[-2\phi_{1,2}-\phi_{2,1}]$ and $[\phi_1-\phi_2]$ divide the blue and yellow domains both into six new domains, and the unrefined points lie at the intersection of these domains.}
	\label{fig:finite-n-domain}
\end{figure}

At finite $N$, we should take into account all terms in \eqref{eq:Qtot-prime-suN}.
Explicitly, we have:
\begin{align}\label{eq:Qtot-prime-suN-finiteN}
\begin{split}
Q_{\mathrm{tot}}(\phi_{a_{i}})&= (N-1)\Big(-Q(0)+Q([\phi_1])+Q([\phi_2])+Q([-\phi_1-\phi_2])\Big)\\
&+\sum_{\phi\in\lbrace\phi_{a_i}-\phi_{a_j}\rbrace}\sum^3_{b=1}\Big(Q([\phi+\phi_b])+Q([-\phi+\phi_b]))\Big)\\
&+\sum_{\phi\in\lbrace\phi_{a_i}\rbrace}\sum^3_{b=1}\Big(Q([\phi+\phi_b])+Q([-\phi+\phi_b])\Big).
\end{split}
\end{align}
The first line gives the expected entropy function in the blue domain of Figure \ref{fig:large-n-domain}, while the second line will gives us the entropy function up to remainder terms proportional to $(\phi_{a_i}-\phi_{a_j})^2$ as we explained above. 
However, the last line gives rise to new brackets and remainder terms.
There are again three possible terms:
\begin{align}\label{eq:Qs-phia}
\begin{split}
Q_{1}&= Q([2\phi_1])+Q([\phi_1+\phi_2])+Q([-\phi_1+\phi_2])+Q([-\phi_2])+Q([-2\phi_1-\phi_2])\\
Q_{2}&= Q([\phi_2+\phi_1])+Q([-\phi_2+\phi_1])+Q([2\phi_2])+Q([-\phi_1])+Q([-2\phi_2-\phi_1])\\
Q_{3}&= Q([-\phi_2])+Q([-2\phi_1-\phi_2])+Q([-\phi_1])+Q([-2\phi_2-\phi_1])\\
&+Q([-2\phi_1-2\phi_2]).
\end{split}
\end{align}
At large-$N$ we have been able to ignore the fact that our computation of the index really requires us to stay away from the unrefined point, since the $\mathcal{O}(N^2)$ part of the $Q^\prime_{\mathrm{tot}}$ remains unchanged in the proper calculation (see Appendix \ref{app:unrefined-limit-index}).
However, the $Q$ functions in \eqref{eq:Qs-phia} coming in at $\mathcal{O}(N)$ do not appear in the unrefined limit and should therefore not be evaluated at the unrefined point.
This can also be seen from the terms with the brackets $[-2\phi_{1,2}-\phi_{2,1}]$ and $[\phi_1-\phi_2]$, which are not defined at the unrefined points.
In particular, these brackets divide the blue and yellow domain of Figure \ref{fig:large-n-domain} into six new regions.
The unrefined point lies precisely at the intersection of the boundaries of regions.
See Figure \ref{fig:finite-n-domain}. 

For this reason, we need to use the results from Appendix \ref{app:unrefined-limit-index} to study the Cardy limit of the index at finite $N$ at the unrefined points.
The finite $N$ $Q$-polynomials are given in \eqref{eq:Q-pols-unref-finite-N}, which we repeat here for convenience:
\begin{align}
\begin{split}
Q_1&= \frac{N^2-3N+2}{9\tau\sigma}+\mathcal{O}\left(\tau^{-1}\right)+\mathcal{O}\left(\sigma^{-1}\right),\\
Q_2&=  -\frac{N^2-3N+2}{9\tau\sigma}+\mathcal{O}\left(\tau^{-1}\right)+\mathcal{O}\left(\sigma^{-1}\right).
\end{split}
\end{align}
These functions correspond to the total $Q$ polynomials at the unrefined point in the blue and yellow domain respectively.
The fact that we do not see the expected $N^2-1$ is explained at the end of Appendix \ref{app:suN-index}.

At finite $N$, we cannot move away as nicely from the unrefined point as we did at large-$N$.
The reason for this is that the unrefined point lies at the intersection of the domains in the refined computation.
Because in each of the six domains, the set of brackets takes on a different value, moving away from the unrefined point in this case depends on the direction one is going in. 
In particular, the resulting remainder terms associated to any of the domains are not small because of the discontinuous nature of the brackets.

\section{Discussion}\label{sec:conc}

In this paper, we have computed the superconformal index of the $\mathcal{N}=4$ theory through residues.
This method is the same as the one used to derive the Higgs branch localization formula from the index, which as far as we are aware was only applied in the literature to gauge theories with fundamental matter (see e.g.\ \cite{Yoshida:2014qwa,Peelaers:2014ima,Nieri:2015yia} and references therein).\footnote{In the context of the Hilbert series, similar residue computations were performed for SQCD in \cite{Gray:2008yu} and for adjoint SQCD in \cite{Hanany:2008sb}.}
We have seen that in the refined computation, the adjoint matter results in non-degenerate non-factorized poles.
Because the poles are non-degenerate, it is still fairly easy to compute the residue.\footnote{In Appendix \ref{app:unrefined-limit-index}, we compute the index in the unrefined limit. In this case, the poles are degenerate and one needs more sophisticated techniques to compute the residue, as we discuss in detail.}
The real complication shows up in deciding when a pole falls inside all unit circles, which depends on the precise values of the chemical potentials.
This prohibits us from finding a fully explicit expression for the index at general values of the parameters.
Luckily, for purposes of taking the Cardy limit we have been able to argue that this complication is irrelevant.
Studying the Cardy limit of our expression, we first of all find that all residues contribute at leading order.
In addition, close to the unflavoured or unrefined points in chemical potential space, all residues become equal to the entropy function of the 1/16$^{\mathrm{th}}$ BPS asymptotically AdS$_5$ black hole and its ``twin saddle'' respectively (see the remark in footnote \ref{fn:ofer}).

The Bethe Ansatz method, as developed in \cite{Closset:2017bse,Closset:2018ghr,Benini:2018mlo}, also computes the gauge integral through residues and was applied to compute the large-$N$ limit and the study of AdS$_5$ black holes in \cite{Benini:2018ywd,Benini:2020gjh}.
However, our method is technically distinct from the Bethe Ansatz method and the final formulas have important differences.
Let us discuss this in some detail.
Firstly, the following rewriting of the gauge integrand is employed in \cite{Benini:2018mlo}:
\begin{align}\label{eq:integrand}
\begin{split}
I(p,q,f)&=\frac{\kappa_N}{N!}\prod^{N-1}_{i=1}\oint_{\left|x_i\right|=1}\frac{dx_i}{2\pi i x_i}\mathcal{Z}(u;\phi,a\omega,b\omega)\\
&=\frac{\kappa_N}{N!}\oint_{\mathcal{C}}\prod^{N-1}_{i=1}\frac{dx_i}{2\pi i x_i}\frac{\mathcal{Z} (u;\phi,a\omega,b\omega)}{\prod_{i=1}^{N-1} (1-\tilde{Q}_i)}
\end{split}
\end{align}
where $\mathcal{Z}$ is short-hand notation for the integrand (see e.g.\ \eqref{eq:defn-suN-index}).
The chemical potentials coupling to the angular momenta are specialized: $\tau=a \omega$ and $\sigma= b \omega$ with $a,b\in \mathbb{Z}$. 
Furthermore, $\mathcal{C}$ is a new contour and $\tilde{Q}_i=1$ represent the so-called Bethe Ansatz equations. 
Inside this new contour, one only has to take into account poles coming from the denominator, i.e.\ solutions to the Bethe Ansatz equations.
This is a very different set of poles than the set we consider, which originate from $\mathcal{Z}$.
In particular, the basic solutions to the Bethe Ansatz equations do not depend on the R-symmetry chemical potentials $\phi_a$, whereas all our poles do.\footnote{See however \cite{ArabiArdehali:2019orz} for non-standard solutions to the Bethe Ansatz equations that do depend on the $\phi_a$.}
Moreover, in our case all poles are explicitly known even though whether the associated residue contributes depends on the precise values of the chemical potentials.
In contrast, in the Bethe Ansatz approach not all poles are known due to difficulties in solving the Bethe Ansatz equation.
However, in their case the residue associated to a pole will always contribute.
Another difference that stands out is that in the Bethe Ansatz method at large-$N$, there is generically a single dominant residue which captures the correct entropy functions.
For us instead, it seems that generically all residues contribute at leading order in the Cardy limit, as we have discussed at length in Section \ref{ssec:cardy-limit-index}.
It would be very interesting to compare the two methods.
Since we only have a fully explicit general expression for $SU(2)$ gauge group, this may be a good place to start.
Otherwise, for $SU(N)$ one would have to consider a specific regime in parameter space where our expression can be made fully explicit.
We leave this as future work.

This work was originally motivated to understand what role certain modular properties of four-dimensional supersymmetric partition functions \cite{Gadde:2020bov} play in the evaluation of the $\mathcal{N}=4$ $SU(N)$ superconformal index and the associated gravitational interpretation.
Indeed, a very interesting question is whether in AdS$_5$/CFT$_4$ there is an analogous story to the Farey tail expansion of the elliptic genus and the $SL(2,\mathbb{Z})$ family of BTZ black holes familiar from AdS$_3$/CFT$_2$ \cite{Maldacena:1998bw,Dijkgraaf:2000fq,Manschot:2007ha}.
Such a story may also connect to the $(m,n)$ saddles of the matrix integral found in \cite{Cabo-Bizet:2019eaf} using an elliptic extension of the gauge integrand.
We believe our expression for the index is suited for such a study for the following basic reason: we have performed the gauge integral before taking the Cardy limit, which firstly allows us to justify the use of the modular property a priori and secondly makes the relation between modularity and the Cardy limit more transparent.
Instead, taking the Cardy limit at the level of the gauge integrand as in e.g.\ \cite{Choi:2018hmj,Honda:2019cio,ArabiArdehali:2019tdm}, these two consequences are less transparent.
Given our expression, we can study more general modular properties obeyed by the elliptic $\Gamma$ function \cite{Gadde:2020bov}.
Even though the modular property used in this paper is suited for the study of the Cardy limit $\tau,\sigma\to 0$, such other modular properties can be useful to study more general ``Cardy'' limits.
In addition, they can help to explore the phase structure suggested by \cite{Cabo-Bizet:2019eaf} within our formalism.
This will be the subject of a future publication \cite{Jejjala:2021hlt}.

In a previous work \cite{Goldstein:2019gpz} we studied a CFT$_2$ subsector of the $\mathcal{N}=4$ theory from the perspective of the superconformal index.
We found that a two-dimensional Cardy formula arises from the superconformal index, which is indicative of ordinary $SL(2,\mathbb{Z})$ modularity.
It would be interesting to understand how this relates to the modularity mentioned in the previous paragraph.
See also \cite{Razamat:2012uv} for earlier and possibly related work. 
Other future directions include considering $\mathcal{N}=1$ superconformal field theories, which have been studied in the Cardy limit in \cite{Cabo-Bizet:2019osg,Kim:2019yrz}, by Bethe Ansatz methods in \cite{Lanir:2019abx,Lezcano:2019pae,Benini:2020gjh}, and finally by large-$N$ saddle point approximation in \cite{Cabo-Bizet:2020nkr}.
Finally, it would be interesting to understand if our expression allows one to compute subleading corrections in the Cardy limit.

\section*{Acknowledgements}

We are grateful to Yang Zhang for useful discussions. 
We also would like to thank the organizers and speakers of the online workshop ``Supersymmetric Black Holes, Holography and Microstate Counting'' for providing us with excellent talks and discussion sessions.
KG, VJ, and SvL are supported by the Simons Foundation Mathematical and Physical Sciences Targeted Grants to Institutes, Award ID:509116.
VJ is supported by the South African Research Chairs Initiative of the Department of Science and Technology and the National Research Foundation.  
YL is supported by the South African Research Chairs Initiative of the Department of Science and Technology and the National Research Foundation, also in part by the project “Towards a deeper understanding of black holes with non-relativistic holography” of the Independent Research Fund Denmark (grant number DFF-6108-00340) and by the UCAS program of special research associate and by the internal funds of the KITS. 
WL is supported by NSFC No.\ 11875064, No.\ 11947302, and the Max-Planck Partergruppen fund.

\appendix

\section{Properties of \texorpdfstring{$\theta$}{theta} and elliptic \texorpdfstring{$\Gamma$}{Gamma} functions}\label{app:props-theta-gamma}

We collect here the definitions and some important properties of the $\theta_q$ and the elliptic $\Gamma$ functions, which are used in the main text.
We have taken most formulae from the work \cite{Felder_2000}.

\paragraph{$\theta$ function:}

The $\theta_q$ function, also known as the $q$-theta function, can be defined as an infinite product for $\mathrm{Im}(\tau)>0$:
\begin{equation}\label{eq:defn-theta}
\theta_q(x)\equiv\theta(z;\tau)=(x;q)_{\infty}\:(qx^{-1};q)_{\infty}=\prod^{\infty}_{n=0}(1-xq^{n})(1-x^{-1}q^{n+1}).
\end{equation}
where $q=e^{2\pi i \tau}$, $x=e^{2\pi i z}$ and the $q$-Pochhammer symbol is defined as:
\begin{equation}\label{eq:defn-qpoch}
(x;q)_{\infty}=\prod^{\infty}_{n=0}(1-xq^n).
\end{equation}
Alternatively, there is the summation formula defined for $0<\mathrm{Im}(z)<\mathrm{Im}(\tau)$:
\begin{equation}\label{eq:theta-sum-form}
\theta(z;\tau)=\exp\left(-\sum^{\infty}_{l=1}\frac{1}{l}\frac{x^l+(qx^{-1})^l}{(1-q^l)}\right)
\end{equation}
The $\theta$ function is quasi-elliptic under the translation $z\to z+m\tau+n$, $m,n\in \mathbb{Z}$:
\begin{equation}\label{eq:shift-prop-theta}
\theta(z+m\tau+n;\tau)=(-x)^{-m}q^{-\frac{m(m-1)}{2}}\theta(z;\tau).
\end{equation}
Furthermore, it satisfies a reflection property:
\begin{equation}
\theta(-z;\tau)=\theta(z+\tau;\tau)=-x^{-1}\theta(z;\tau),
\end{equation}
and can be extended to $\mathrm{Im}(\tau)<0$ through:
\begin{equation}\label{eq:theta-extend}
\theta(z;-\tau)\equiv \frac{-x}{\theta(z;\tau)}.
\end{equation}
Finally, the $\theta$ function satisfies a modular property:
\begin{equation}\label{eq:mod-prop-theta-app}
\theta(z;\tau)=e^{-i\pi B(z,\tau)}\theta\left(\frac{z}{\tau};-\frac{1}{\tau}\right),
\end{equation}
where:
\begin{equation}\label{eq:B-pol}
B(z,\tau)=\frac{z^2}{\tau}+z\left(\frac{1}{\tau}-1\right)+\frac{1}{6}\left(\tau+\frac{1}{\tau}\right)-\frac{1}{2}.
\end{equation}
Another version of the modular property, which we also require, is given by:
\begin{equation}\label{eq:mod-prop-theta-2}
\theta\left(\frac{z}{\tau};\frac{\sigma}{\tau}\right)\theta\left(\frac{z}{\sigma};\frac{\tau}{\sigma}\right)=e^{-i\pi\Phi(z,\tau,\sigma)},
\end{equation}
where:
\begin{equation}\label{eq:defn-Phi}
\Phi(z,\tau,\sigma)=\frac{1}{\tau\sigma}\left(\left(z-\frac{\tau+\sigma}{2}\right)^2-\frac{\tau^2+\sigma^2}{12}\right).
\end{equation}

\paragraph{Elliptic $\Gamma$ function:}

The elliptic $\Gamma$ function can be defined as an infinite product when $\mathrm{Im}(\tau),\mathrm{Im}(\sigma)>0$ as follows:
\begin{align}\label{eq:defn-ell-gamma-app}
\begin{split}
\Gamma(x)\equiv\Gamma(z;\tau,\sigma)=\prod^{\infty}_{m,n=0}\frac{1-x^{-1}p^{m+1}q^{n+1}}{1-x p^{m}q^{n}},
\end{split}
\end{align}
where $q=e^{2\pi i \tau}$, $p=e^{2\pi i \sigma}$ and $x=e^{2\pi i z}$.
Alternatively, for $\mathrm{Im}(\tau),\mathrm{Im}(\sigma)>0$ and $0<\mathrm{Im}(u)<\mathrm{Im}(\tau)+\mathrm{Im}(\sigma)$ it can also be defined through the summation formula:
\begin{equation}\label{eq:defn-ell-gamma-sum}
\Gamma(z;\tau,\sigma)=\exp\left(\sum^{\infty}_{l=1}\frac{1}{l}\frac{x^l-(x^{-1}pq)^l}{(1-p^l)(1-q^l)}\right),
\end{equation}
Basic properties that are manifest from these expressions include:
\begin{align}
\begin{split}
\Gamma(z+1;\tau,\sigma)&=\Gamma(z;\tau+1,\sigma)=\Gamma(z;\tau,\sigma+1)=\Gamma(z;\tau,\sigma),\\
\Gamma(z;\tau,\sigma)&=\Gamma(z;\sigma,\tau).
\end{split}
\end{align} 
Furthermore, the elliptic $\Gamma$ function satisfies the following basic shift properties:
\begin{align}\label{eq:basic-shift-gamma-app}
\begin{split}
\Gamma(z+\tau;\tau,\sigma)&=\theta(z;\sigma)\Gamma(z;\tau,\sigma)\\
\Gamma(z+\sigma;\tau,\sigma)&=\theta(z;\tau)\Gamma(z;\tau,\sigma).
\end{split}
\end{align}
Repetitive use of these properties and the shift property of the $\theta$ function \eqref{eq:shift-prop-theta} leads to:
\begin{align}\label{eq:shift-props-gammas-app}
\begin{split}
\Gamma(p^{k}q^{l}x)&=\Gamma(x)\left(-xp^{\frac{k-1}{2}}q^{\frac{l-1}{2}}\right)^{-kl}\prod^{k-1}_{m=0}\theta_q(xp^m)\prod^{l-1}_{n=0}\theta_p(xq^n),\\
\Gamma(p^{k}q^{-l}x)&=\Gamma(x)\frac{\prod^{k-1}_{m=0}\theta_q(xp^{-m})}{\left(-x^{-1}p^{\frac{k-1}{2}}q^{-\frac{l+1}{2}}\right)^{-kl}\prod^{l}_{n=1}\theta_p(xq^{-n})}\\
\Gamma(p^{-k}q^{l}x)&=\Gamma(x)\frac{\prod^{l-1}_{n=0}\theta_p(xq^{-n})}{\left(-x^{-1}p^{-\frac{k+1}{2}}q^{\frac{l-1}{2}}\right)^{-kl}\prod^{k}_{m=1}\theta_q(xp^{-m})}\\
\Gamma(p^{-k}q^{-l}x)&=\Gamma(x)\frac{1}{\left(-x^{-1}p^{\frac{k+1}{2}}q^{\frac{l+1}{2}}\right)^{-kl}\prod^{k}_{m=1}\theta_q(xp^{-m})\prod^{l}_{n=1}\theta_p(xq^{-n})},
\end{split}
\end{align}
where for convenience we use the shorthand notation for $\theta$ and $\Gamma$.
Moreover, for $k=0$ ($l=0$) the product over $m$ ($n$) is defined to be $1$.

Another important property we use is given by:
\begin{equation}\label{eq:gamma-theta-id-app}
\Gamma(z;\tau,\sigma)\Gamma(-z;\tau,\sigma)=\frac{1}{\theta(z;\sigma)\theta(-z;\tau)}.
\end{equation}

The elliptic $\Gamma$ function can be extended to the lower half planes $\mathrm{Im}(\tau)<0$ or $\mathrm{Im}(\sigma)<0$ via the summation formula \eqref{eq:defn-ell-gamma-sum}.
Specifically, we have: 
\begin{align}\label{eq:extend}
\begin{split}
& \Gamma(z;-\tau,\sigma) = \frac{1}{\Gamma(z+\tau;\tau,\sigma)} = \Gamma(\sigma-z;\tau,\sigma) \\
& \Gamma(z;\tau,-\sigma) = \frac{1}{\Gamma(z+\sigma;\tau,\sigma)} = \Gamma(\tau-z;\tau,\sigma) 
\end{split}
\end{align}
With these expressions in mind, the elliptic Gamma function is defined for $\tau,\sigma \in \mathbb{C}-\mathbb{R}$.

Finally, the elliptic $\Gamma$ function satisfies a modular property, which given \eqref{eq:extend} is defined for $\tau,\sigma,\frac{\tau}{\sigma}\in \mathbb{C}\setminus \mathbb{R}$:
\begin{equation}\label{eq:mod-prop-Gamma-app}
\Gamma(z;\tau,\sigma)=e^{i\frac{\pi}{3} Q(z,\tau,\sigma)}\frac{\Gamma\left(\frac{z}{\tau};\frac{\sigma}{\tau},-\frac{1}{\tau}\right)}{\Gamma\left(\frac{z-\tau}{\sigma};-\frac{\tau}{\sigma},-\frac{1}{\sigma}\right)}.
\end{equation}
Here, $Q(z,\tau,\sigma)$ is the following polynomial:
\begin{equation}
Q(z,\tau,\sigma)=\frac{-1}{\tau\sigma}\left(z-\frac{\tau+\sigma}{2}+\frac{1}{2}\right)\left(\left(z-\frac{\tau+\sigma}{2}\right)^2+\left(z-\frac{\tau+\sigma}{2}\right)-\frac{\tau^2+\sigma^2}{4}\right).
\end{equation}
This is the same modular property as equation (21) of Theorem 4.1 in \cite{Felder_2000}, where our $Q$ polynomial is related to theirs by: $Q_{\mathrm{ours}}=-3Q_{\mathrm{theirs}}$.

We point out that a slightly different version of the modular property is used in \cite{Gadde:2020bov}:
\begin{equation}\label{eq:mod-prop-Gamma-Gadde}
\Gamma(z;\tau,\sigma)\Gamma\left(\frac{z}{\tau};\frac{\sigma}{\tau},\frac{1}{\tau}\right)\Gamma\left(\frac{z}{\sigma};\frac{1}{\sigma},\frac{\tau}{\sigma}\right)=e^{-i\frac{\pi}{3} Q_G(z,\tau,\sigma)},
\end{equation}
with:
\begin{align}
\begin{split}
Q_G(z,\tau,\sigma)&=\frac{z^3}{\tau\sigma}-\frac{3}{2}\frac{\tau+\sigma+1}{\tau\sigma}z^2+\frac{\tau^2+\sigma^2+3\tau \sigma+3\tau+3\sigma+1}{2\tau\sigma}z\\
&-\frac{1}{4}\left(\tau+\sigma+1\right)\left(\frac{1}{\tau}+\frac{1}{\sigma}+1\right).
\end{split}
\end{align}
We can rewrite this property into ours by making use of the formulas given above.
In particular, we first use the extension formulas \eqref{eq:extend} and the shift properties \eqref{eq:shift-props-gammas-app} of the elliptic $\Gamma$ function.
Subsequently, we use the shift property \eqref{eq:shift-prop-theta} and the extension formula \eqref{eq:theta-extend} for the $\theta$ function, and finally the second modular property of the $\theta$ function \eqref{eq:mod-prop-theta-2}.
Doing this, one finds our modular property, where:
\begin{equation}
Q(z,\tau,\sigma)=-Q_G(z,\tau,\sigma)-3\Phi(z,\tau,\sigma).
\end{equation}
The reason we choose to use \eqref{eq:mod-prop-Gamma-app} is that the product expressions for the given arguments manifestly converge when $\mathrm{Im}(\tau),\mathrm{Im}(\sigma),\mathrm{Im}\left(\frac{\sigma}{\tau}\right)>0$.

\section{Unrefined limit of the index}\label{app:unrefined-limit-index}

In this appendix, we consider the unrefined limit of the index: 
\begin{equation}
f\equiv f_1=f_2=f_3=(pq)^{\frac{1}{3}}.
\end{equation}
This limit has to be treated separately because any set of three simple poles $y=f_ap^kq^l$ associated to three $\Gamma$ factors of the form $\prod^3_{b=1}\Gamma(yf_b)$ collide into a single cubic pole.
Even though the final expression for the index is different in some respects, we will find that, at large-$N$, the leading order expression in the Cardy limit remains unchanged.
This justifies the naive unrefined limit of the Cardy limit of the refined index computed in the main text.
As in the above, we treat $SU(2)$ and $SU(N)$ separately.
 
\subsection{\texorpdfstring{$SU(2)$}{SU(2)} index}\label{app:su2-index}

The expression for the index of the $\mathcal{N}=4$ $SU(2)$ theory in the unrefined limit reads: 
\begin{equation}
I_2 = \frac{\kappa_2}{2} \oint_{|x|=1} \frac{dx}{2\pi i x} \frac{\Gamma(x^{ 2} f)^3 \Gamma(x^{-2} f)^3}{\Gamma(x^2)\Gamma(x^{-2})}
\end{equation}
Inside the unit circle, there are now cubic poles at:
\begin{equation}
x^{2}=f p^{k}q^{l},
\end{equation}
for $k,l\geq 0$.

Let us first compute the residue of $\Gamma(x)^3$ at its basic cubic pole, $x=1$.
We first write:
\begin{equation}
\Gamma(x)^3=\frac{1}{(1-x)^3}\left(\prod_{m,n\geq 0}\frac{1-x^{-1}p^{m+1}q^{n+1}}{1-xp^{m+1}q^{n+1}}\times \prod_{m\geq 0}\frac{1}{1-xp^{m+1}}\frac{1}{1-xq^{m+1}}\right)^3.
\end{equation}
To obtain the residue, Cauchy's integral formula tells us that we have to compute the second derivative of the function in brackets and evaluate it at $x=1$.
Let us call this function $\widehat{\Gamma}(x)$.
We first rewrite $\widehat{\Gamma}(x)$ using the plethystic exponential (cf. \eqref{eq:defn-ell-gamma}):
\begin{align}\label{eq:gamma-hat}
\begin{split}
\widehat{\Gamma}(x)=e^{\gamma(x)}=\exp\left[\sum^{\infty}_{l=1}\frac{1}{l}\left(\frac{-(xpq)^l-(x^{-1}pq)^l+x^l(p^l+q^l)}{(1-p^l)(1-q^l)}\right)\right].
\end{split}
\end{align}
The residue can be written in terms of $\gamma$ as follows: 
\begin{equation}\label{eq:res-basic-cubic-pole-Gamma}
 \text{Res}_{x=1} \left( \frac{\widehat{\Gamma}(x)^3}{(1-x)^3} \right) = -\tfrac{3}{2} e^{3\gamma(1)} \left(3\gamma'(1)^2+\gamma''(1)\right)
\end{equation}
Here, we have:
\begin{align}\label{eq:log-gamma-derivs}
\begin{split}
\gamma'(x) &= \sum_{l=1}^{\infty}  \frac{-(pq)^l(x^{l-1}-x^{-l-1})+x^{l-1}(p^l+q^l)}{(1-p^l)(1-q^l)}, \\
\gamma''(x) &= \sum_{l=1}^\infty \frac{(l-1)x^{l-2}(p^l+q^l) - (l-1)(pq)^l x^{l-2} -(l+1)(pq)^l x^{-l-2}}{(1-p^l)(1-q)^l}.
\end{split}
\end{align}
This leads us to our final expression for the residue at the cubic pole:
\begin{align}
\begin{split}
 \text{Res}_{x=1}& \left( \frac{\widehat{\Gamma}(x)^3}{(1-x)^3} \right)=-\frac{3}{2}\frac{1}{(p;p)_{\infty}^3(q;q)_{\infty}^3}\\
 &\times\left[3\left(\sum_{l=1}^{\infty} \frac{p^l+q^l}{(1-p^l)(1-q^l)}\right)^2+\sum_{l=1}^\infty \frac{(l-1)(p^l+q^l)-2l (pq)^l}{(1-p^l)(1-q^l)}\right].
\end{split}
\end{align}
This residue, together with the use of shift properties of the elliptic $\Gamma$ functions as in Section \ref{ssec:su2-index}, allows us to find for the unrefined $SU(2)$ index:
\begin{align}\label{eq:su2-index-final-unrefined}
\begin{split}
&I_2=-\frac{3\Gamma(f)^3\left(3\gamma'(1)^2+\gamma''(1)\right)}{4(p;p)_{\infty}^2(q;q)_{\infty}^2}\frac{\Gamma(f^2)^3}{\Gamma(f)\Gamma(f^{-1})}Z_V(\phi,\sigma;\tau)Z_V(\phi,\tau;\sigma),
\end{split}
\end{align} 
where:
\begin{align}\label{eq:vortex-factors-unrefined}
\begin{split}
Z_{V}(\phi,\sigma;\tau)= \sum_{k\geq 0}\frac{\prod^k_{m=1}\theta_q(f^{-1}p^{-m})}{\prod^{k-1}_{m=0}\theta_q(fp^{m})}\left(\frac{\prod^{k-1}_{m=0}\theta_q(f^2p^{m})}{\prod^k_{m=1}\theta_q(p^{-m})}\right)^3.
\end{split}
\end{align} 

\subsection{\texorpdfstring{$SU(N)$}{SU(N)} index}\label{app:suN-index}

We will now attempt a similar computation for the $SU(N)$ theory.
The unrefined limit of the index is given by:
\begin{align}\label{eq:sci-suN-unrefined}
\begin{split}
I_N=\frac{\kappa_N}{N!}\prod^{N-1}_{k=1}\oint_{\left|x_k\right|=1}\frac{dx_k}{2\pi i x_k}\prod^{N-1}_{i< j}\frac{\Gamma(x_{ij}^{\pm}f)^3}{\Gamma(x_{ij}^{\pm})}\prod^{N-1}_{i=1}
\frac{\Gamma((x_1 \cdots x_i^2\cdots x_{N-1})^{\pm }f)^3}{\Gamma((x_1 \cdots x_i^2\cdots x_{N-1})^{\pm })}.
\end{split}
\end{align}
In this case, poles arise from the intersection of any solvable set of $N-1$ of the following $N^2-N$ equations:
\begin{align}\label{eq:lin-eqns-set-unrefined}
\begin{split}
&u_i-u_j=\phi+k_{ij}\sigma+l_{ij}\tau, \qquad 1\leq i \neq j\leq N-1,\\
&u_1+\ldots  +2u_i+\ldots+u_{N-1}=\mp (\phi+k_{i}\sigma+l_{i}\tau), \qquad i=1,\ldots,N-1,
\end{split}
\end{align}
where we defined $f=e^{2\pi i \phi}$.

For the same reasons as in Section \ref{ssec:suN-index}, we only consider poles originating from the pole equations on the second line of \eqref{eq:lin-eqns-set-unrefined} with all $+$ signs.
However, in contrast with the analysis in Section \ref{ssec:suN-index}, it is not possible to use the formula \eqref{eq:non-deg-res} for a multivariate residue.
We will now briefly discuss the reason for this and provide an alternative formula that can be used.

Firstly, given the $N-1$ equations of the second type with all $+$ signs, a natural choice for the functions $g_i(x)$ is:\footnote{Other choices are formed by choosing different distributions of the singular factors over the $g_i$. However, our choice is singled out because it is the unique choice (up to relabelling) that respects the $S_{N-1}$ symmetry of the integral.}
\begin{equation}\label{eq:gi-deg}
g_i(x)=x_i\left(1-\frac{fp^{k_i}q^{l_i}}{x_1\cdots x_i^2\cdots x_{N-1}}\right)^3.
\end{equation}
The issue with higher order poles is that in this case:
\begin{equation}
J_{p^{(n)}}=\det\left(  \frac{\partial g_i}{\partial x_j}  \right)\bigg|_{x = p^{(n)}}=0.
\end{equation}
This prohibits the use of the formula \eqref{eq:non-deg-res} to compute the residue, which is only defined for non-degenerate residues with $J_p\neq 0$.

Luckily, there exist more sophisticated techniques to compute the residue of such degenerate multivariate residues.
The main formula for the degenerate case is reviewed around Theorem 1 in \cite{Larsen:2017aqb}, in which also additional references may be found.
The basic idea is still similar to the non-degenerate case: one wants to find a transformation that factorizes the multivariate residue integral into a product of univariate ones.
The main formula, Theorem 1 in \cite{Larsen:2017aqb}, for the evaluation of degenerate multivariate residues is given by:
\begin{align}\label{eq:deg-res}
\mathrm{Res}_{x=p}\left(\frac{h(x) d x_1\cdots d x_n}{g_1(x)\cdots g_n(x)}\right)
=\mathrm{Res}_{x=p}\left(\frac{h(x)\det A(x) d x_1\cdots d x_n}{g^{\prime}_1(x_1)\cdots g^\prime_n(x_n)}\right) .
\end{align}
This formula is the analogue of \eqref{eq:non-deg-res} for a degenerate pole, i.e.\ an isolated zero at $x=p$ of $g(x)=(g_1(x),\ldots,g_n(x))$ with $J_p=0$.
In this formula, the $g^{\prime}_i(x_i)$ are functions that only depend on $x_i$, and can be obtained from the $g_i(x)$ via:
\begin{equation}
g^{\prime}_i(x_i)=\sum_ja_{ij}(x)g_j(x),
\end{equation}
where $A(x)=(a_{ij}(x))$ is a matrix of holomorphic polynomials.
The polynomial $A(x)$ can be determined algorithmically through a so-called Gr\"obner basis computation. 
For details, we again refer to \cite{Larsen:2017aqb} and references therein.

We have done the Gr\"obner basis calculation through Mathematica at low values of $N$.
These computations suggest that the $g_i^\prime$ for general $N$ are given by:
\begin{equation}
g_i^\prime(x_i)=(x_i^N-f)^{2N-1},
\end{equation}
where we have chosen to only to the computation for $k_i,l_i=0$.
We do not have to understand the more general case, since any $p,q$ shifted pole can be first brought back to this basic pole with the help of shift properties of the elliptic $\Gamma$ functions.
The formula for $g_i^\prime(x_i)$ shows that the perhaps from \eqref{eq:gi-deg} naively expected cubic poles are actually poles of order $2N-1$.

We now continue to compute the residue at the pole (cf. \eqref{eq:pole-suN}):
\begin{equation}\label{eq:pole-suN-unrefined}
p^{(n)}: \qquad x_i^N=fp^{Nk_i-\sum_{j}k_j}q^{Nl_i-\sum_{j}l_j}.
\end{equation}
Firstly, we bring the residue back to $x_i^N=f$ using the shift properties of the elliptic $\Gamma$ function:
\begin{align}\label{eq:residue-basic-pole-suN-unrefined}
\begin{split}
&\mathrm{Res}_{\sum p^{(n)}}\left(\frac{1}{x_1x_2\cdots x_{N-1}}\prod^{N-1}_{i< j}\frac{\Gamma(x_{ij}^{\pm}f)^3}{\Gamma(x_{ij}^{\pm})}\prod^{N-1}_{i=1}
\frac{\Gamma((x_1 \cdots x_i^2\cdots x_{N-1})^{\pm }f)^3}{\Gamma((x_1 \cdots x_i^2\cdots x_{N-1})^{\pm })}\right)\\
&=\prod^{N-1}_{i< j}\frac{\Gamma((p^{k_i-k_j}q^{l_i-l_j})^{\pm}f)^3}{\Gamma((p^{k_i-k_j}q^{l_i-l_j})^{\pm})}\prod^{N-1}_{i=1}
\frac{\Gamma(f^2 p^{k_i}q^{l_i})^3}{\Gamma((f p^{k_i}q^{l_i})^{\pm })}\\
&\times\mathrm{Res}_{\sum p^{(n)}}\left(\prod^{N-1}_{i=1}\frac{\Gamma((x_1 \cdots x_i^2\cdots x_{N-1})^{-1 }f)^3}{x_i}\right)\\
&=\prod^{N-1}_{i< j}\frac{\Gamma((p^{k_i-k_j}q^{l_i-l_j})^{\pm}f)^3}{\Gamma((p^{k_i-k_j}q^{l_i-l_j})^{\pm})}\prod^{N-1}_{i=1}
\frac{\Gamma(f^2 p^{k_i}q^{l_i})^3}{\Gamma((f p^{k_i}q^{l_i})^{\pm })} \left(\frac{1}{C_i\prod^{k_i}_{m=1}\theta_q(p^{-m})\prod^{l_i}_{n=1}\theta_p(q^{-n})}\right)^3\\
&\times  \mathrm{Res}_{x_i^N=f}\left(\prod^{N-1}_{i=1}\frac{\Gamma((x_1 \cdots x_i^2\cdots x_{N-1})^{-1 }f)^3}{x_i}\right),
\end{split}
\end{align}
where:
\begin{equation}
C_i=\left(-p^{\frac{k_i+1}{2}}q^{\frac{l_i+1}{2}}\right)^{-k_il_i}.
\end{equation}
Now, we use the formula \eqref{eq:deg-res} for the last line:
\begin{align}
\begin{split}
&\mathrm{Res}_{x_i^N=f}\left(\prod^{N-1}_{i=1}\frac{\Gamma((x_1 \cdots x_i^2\cdots x_{N-1})^{-1 }f)^3}{x_i}\right)\\
&=\prod^{N-1}_{i=1}\mathrm{Res}\left(\frac{\det A(x)\; \widehat{\Gamma}((x_1 \cdots x_i^2\cdots x_{N-1})^{-1 }f)^3}{(x_i^N-f)^{2N-1}}\right)
\end{split}
\end{align}
We will not find an explicit expression for the residue.
This is because of a technical reason: we have been able to find the matrix $A(x)$ for low values of $N$, but it is a somewhat complicated polynomial whose generalization to arbitrary $N$ is not clear to us.
Luckily, for purposes of the Cardy limit, as we will discuss in more detail in the next section, the form of $A(x)$ is unimportant.
We will also see in the next section that the high order of the poles and the associated derivatives will not be of relevance to the Cardy limit either.
Therefore, we hide all these details in a function $R(f,q,p)$.
Then, analogously to the refined analysis of Section \ref{ssec:suN-index}, the expression for the index in the unrefined limit takes the form:
\begin{align}\label{eq:suN-index-final-unrefined}
\begin{split}
I_N^{\prime}=\frac{\Gamma(f)^{3(N-1)^2}\Gamma(f^2)^{3(N-1)}R(f,q,p)}{N!\left((p;p)_{\infty}(q;q)_{\infty}\right)^{2(N-1)}}\sum^\prime_{(k_i),(l_i)\geq (0)} Z^{(k_i)}_{V,}(\phi,\sigma;\tau) Z^{(l_i)}_{V}(\phi,\tau;\sigma),
\end{split}
\end{align}
where the Pochhammer symbols originate from the residue similarly to the $SU(2)$ case discussed in the previous section, and details of the $A(x)$ matrix and derivatives of it and $\widehat{\Gamma}$ are hidden in $R$.
As in the refined case, the precise form of the vortex partition functions of the numerator $Z_{V}$ depends on the sign of $k_i-k_j$ and $l_i-l_j$.
For both positive or both negative for all $i<j$, the vortex partition function is given by:
\begin{align}\label{eq:vortex-part-suN-unrefined}
\begin{split}
 Z^{(k_i)}_{V}(\phi,\sigma;\tau)&=\prod^{N-1}_{i< j}\frac{\prod^{k_i-k_j}_{m=1}\theta_q(p^{-m})}{\prod^{k_i-k_j-1}_{m=1}\theta_q(p^m)}\prod^{N-1}_{i=1}\frac{\prod^{k_i}_{m=1}\theta_q(f^{-1}p^{-m})}{\prod^{k_i-1}_{m=1}\theta_q(fp^m)}\\
&\times\Bigg(\prod^{N-1}_{i< j}\frac{\prod^{k_i-k_j-1}_{m=0}\theta_q(fp^m)}{\prod^{k_i-k_j}_{m=1}\theta_q(fp^{-m})}
\prod^{N-1}_{i=1}\frac{\prod^{k_i-1}_{m=0}\theta_q(f^2p^m)}{\prod^{k_i}_{m=1}\theta_q(p^{-m})}\Bigg)^3.
\end{split}
\end{align}
This is just the specialization $\phi_1=\phi_2=\phi_3$ of the refined vortex partition functions \eqref{eq:vortex-part-suN}.
All further comments made there apply here as well, so we will not repeat them.

\subsection{Cardy limit of the unrefined index}\label{app:cardy-limit}

As is clear from the expression \eqref{eq:suN-index-final-unrefined}, the universal part of the residue is simplified significantly in the unrefined limit of the index.
We already know that to leading order in the Cardy limit, the vortex partition functions do not contribute at leading order (see Section \ref{sec:cardy-limit-index}).
We will now argue that also the $q$-Pochhammer symbols and $R$ do not contribute at leading order.

For the $q$-Pochhammer symbols we first notice that:
\begin{equation}
(q;q)_{\infty}=q^{-\frac{1}{24}}\eta(\tau),
\end{equation} 
with $\eta(\tau)$ the Dedekind $\eta$ function, which obeys the modular property:
\begin{equation}
\eta (\tau) = \frac{\eta\left(-\frac{1}{\tau}\right)}{\sqrt{-i\tau}}.
\end{equation}
This implies:
\begin{equation}
\lim_{\tau\to 0^{+i}}\frac{1}{(q;q)_{\infty}}=\lim_{\tau\to 0^{+i}}\frac{\tilde{q}^{-\frac{1}{24}}\sqrt{-i\tau}}{q^{-\frac{1}{24}}(\tilde{q};\tilde{q})_{\infty}}=\mathcal{O}(e^{\frac{2\pi i}{\tau}}),
\end{equation}
where $\tilde{q}=e^{-\frac{2\pi i}{\tau}}$.
This divergence is similar to the $\theta$ functions, of course, and subleading in the Cardy limit.

The function $R(f,q,p)$ consists of a sum of products of derivatives of the polynomial $A(x)$ and derivatives of $\gamma(x)$ (see \eqref{eq:gamma-hat}).
Since $A(x)$ is a finite degree polynomial in $x$, the evaluation of its derivatives on the poles lead to a finite degree polynomial in the fugacities. 
This contributes at order $\mathcal{O}(e^{2\pi i \tau})$ or $\mathcal{O}(e^{2\pi i \sigma})$ and can be safely ignored in the Cardy limit.
For derivatives of $\gamma(x)$ it is not difficult to see, for example from the expressions \eqref{eq:log-gamma-derivs}, they will diverge as $\mathcal{O}(e^{-\log (\tau\sigma)})$, which is again subleading.

We conclude that also in the unrefined case to study the Cardy we only have to consider the $\Gamma$ functions:
\begin{equation}\label{eq:univ-res-unrefined}
\Gamma(f)^{3(N-1)^2}\Gamma(f^2)^{3(N-1)}.
\end{equation}
Since $f=(pq)^{\frac{1}{3}}$ we have: $\phi=\tfrac{1}{3}(\tau+\sigma-k)$ for some $k\in \mathbb{Z}$.
However, evaluation of the brackets, discussed in Section \ref{ssec:mod-prop-theta-gamma}, reduces this choice to the independent values $k=0,1,2$.
It is easy to see that for $k=1,2$ the unrefined points lie inside the diamond $D_0$ in the Cardy limit (see Figure \ref{fig:z-regions-2}).
Instead, for $k=(0\mod 3)$ it will lie outside the diamond.
However, also in this case it is not difficult to see that the Cardy limit of the modular property to leading order only gets a contribution from the $Q$ polynomial.
This is because both $\Gamma$ functions on the right hand side of the modular property \eqref{eq:mod-prop-Gamma}, even though they do not simplify to $1$ in this case, will be convergent functions of $\frac{\sigma}{\tau}$.
Therefore, their contribution will be subleading with respect to the diverging $e^{\frac{i\pi}{3} Q}$ prefactor.

We then proceed to compute the total $Q$ polynomial for these separate cases:
\begin{align}\label{eq:Q-pols-unref-finite-N}
\begin{split}
Q_0&=3(N-1)^2Q([\tfrac{1}{3}(\tau+\sigma)]) +3(N-1)Q([\tfrac{2}{3}(\tau+\sigma)])\\
&=\mathcal{O}\left(\tau^{-1}\right)+\mathcal{O}\left(\sigma^{-1}\right),\\
Q_1&=3(N-1)^2Q([\tfrac{1}{3}(\tau+\sigma-1)]) +3(N-1)Q([\tfrac{2}{3}(\tau+\sigma-1)])\\
&=  \frac{N^2-3N+2}{9\tau\sigma}+\mathcal{O}\left(\tau^{-1}\right)+\mathcal{O}\left(\sigma^{-1}\right),\\
Q_2&=3(N-1)^2Q([\tfrac{1}{3}(\tau+\sigma-2)]) +3(N-1)Q([\tfrac{2}{3}(\tau+\sigma-2)])\\
&=  -\frac{N^2-3N+2}{9\tau\sigma}+\mathcal{O}\left(\tau^{-1}\right)+\mathcal{O}\left(\sigma^{-1}\right).
\end{split}
\end{align}
Here, we have used that the brackets evaluate to leading order as:
\begin{align}
\begin{split}
[\tfrac{1}{3}(\tau+\sigma-1)]&=-\tfrac{1}{3},\qquad [\tfrac{2}{3}(\tau+\sigma-1)]=-\tfrac{2}{3}\\
[\tfrac{1}{3}(\tau+\sigma-2)]&=-\tfrac{2}{3},\qquad [\tfrac{2}{3}(\tau+\sigma-2)]=-\tfrac{1}{3}.
\end{split}
\end{align}
At large-$N$, the expressions for $Q_1$ and $Q_2$ agree with the unrefined limits of the $Q$ polynomials computed away from the unrefined points 
\eqref{eq:Qprime-gen-res} and \eqref{eq:Qprime-gen-res-twin}.
This is what we wanted to show. 
As a final comment, we note that the subleading pieces in $N$ do not take the perhaps expected form of $N^2-1$.
The reason for this is that the missing $3(N-1)$ terms originate from the $q$-Pochhammer symbols, which do not contribute at leading order in the Cardy limit.

\section{Anomaly polynomial}\label{app:anomaly-pol}

Let us parametrize the anomaly polynomial of a general gauge theory as in \cite{Gadde:2020bov}:
\begin{align}
\begin{split}
P(\zeta_a;\omega_i)=\frac{1}{\omega_1\omega_2\omega_3}\left(k_{ijk}\zeta_i\zeta_j\zeta_k+3k_{ijR}\zeta_i\zeta_j\Omega+3k_{iRR}\zeta_i\Omega^2+k_{RRR}\Omega^3-k_i\zeta_i\tilde{\Omega}-k_R\Omega\tilde{\Omega}\right),
\end{split}
\end{align}
where
\begin{equation}
\Omega=\tfrac{1}{2}\sum^3_{i=1}\omega_i,\qquad \tilde{\Omega}=\tfrac{1}{4}\sum^3_{i=1}\omega^2_i.
\end{equation}
We want to compare this general formula to the expression for the anomaly polynomial in terms of the $\phi_a$ \eqref{eq:anom-pol}, i.e.\ :
\begin{equation}
Q_{\mathrm{tot}}(\phi_{a_{i}})=-3(N^2-1)\frac{\phi_1\phi_2\phi_3}{\tau\sigma}.
\end{equation}
with $\phi_3\equiv=\tau+\sigma-\phi_1-\phi_2-1$.
To do this, we set $\omega_1=-1$ and:
\begin{equation}
\tau=\omega_2,\quad\sigma=\omega_3,\quad \phi_a=\zeta_a+\tfrac{2}{3}\Omega.
\end{equation}
Notice that the last definition implies that for $\zeta_a=0$, all $\phi_a$ are equal.
This corresponds to the unrefined limit of the index.
Also, with this identification notice that $\zeta_3=-\zeta_1-\zeta_2$, as expected for the $SU(3)$ flavour fugacities.
We can now express $Q^{a}_{\mathrm{tot}}$ in terms of the new variables:
\begin{equation}
Q^{a}_{\mathrm{tot}}=\frac{3(N^2-1)}{\omega_1\omega_2\omega_3}\left(\zeta_1\zeta_2\zeta_3+\tfrac{2}{3}(\zeta_1\zeta_2+\zeta_1\zeta_3+\zeta_2\zeta_3)\Omega+\tfrac{4}{9}(\zeta_1+\zeta_2+\zeta_3)\Omega^2+\tfrac{8}{27}\Omega^3\right).
\end{equation}
These coefficients encode the global anomalies of the $\mathcal{N}=4$ $SU(N)$ Yang-Mills theory.

\bibliographystyle{JHEP}
\bibliography{bib-bh}

\end{document}